\documentclass[preprint,aps,pra,showpacs,floatfix,superscriptaddress]{revtex4-1}
\usepackage[utf8]{inputenc}
\usepackage[english]{babel}
\usepackage{siunitx}
\usepackage{graphicx}
\usepackage{amsmath}
\usepackage{amsfonts}
\usepackage{amssymb}
\usepackage{amsthm}
\usepackage{epsf}
\usepackage{bm}
\usepackage{color}
\definecolor{BLUE}{rgb}{0.0,0.0,1.0}

\usepackage{soul}
\soulregister\cite7
\soulregister\ref7

\newcommand{\gbb}{g^{(2)}}

\newcommand{\Vm}{V_{\mathrm{magn}}}
\newcommand{\Um}{U_{\mathrm{magn}}}

\newcommand{\gqd}[1]{g^{(2|#1)}}

\newcommand{\dgbbse}{ g^{(2|\mathrm{SE})}}
\newcommand{\dgqse}[1]{g_{#1}^{(2|\mathrm{SE})}}
\newcommand{\dgqvp}[1]{g_{#1}^{(2|\mathrm{VP})}}
\newcommand{\dgbbvp}{ g^{(2|\mathrm{VP})}}

\newcommand{\faz}{F(\alpha Z)}

\newcommand{\ha}{H_1^{\mathrm{rad}}}
\newcommand{\hb}{H_2^{\mathrm{rad}}}

\newcommand{\balpha}{{\mbox{\boldmath$\alpha$}}}
\newcommand{\bnabla}{{\mbox{\boldmath$\nabla$}}}

\newcommand{\be}{\begin{eqnarray}}
\newcommand{\ee}{\end{eqnarray}}

\newcommand{\ket}[1]{\ensuremath{\left|{#1}\right\rangle}}

\newcommand{\mub}{\mu_\textrm{B}}

\newcommand{\Umag}{U_{\textrm{m}}}

\newcommand{\depsn}[1]{(\varepsilon_{a} - \varepsilon_{n_{#1}})}
 
\newcommand{\sumnn}{\sideset{}{'}\sum\limits_{n}}
\newcommand{\summn}{\sideset{}{'}\sum\limits_{n_1,n_2}}

\newcommand{\Vnuc}{V_{\textrm{nuc}}}
\newcommand{\g}{$g$~}

\newcommand{\dgu}{\mbox{$\Gamma'[\Umag](\varepsilon_a)$}}
\newcommand{\gum}{\mbox{$\Gamma[\Umag](\varepsilon_a)$}}
\newcommand{\gumo}{\mbox{$\Gamma[\Umag](\varepsilon)$}}

\newcommand{\lv}{\mbox{$\Lambda(\varepsilon_a)[\Umag,\Umag]$}}
\newcommand{\sige}{{\mbox{$\Sigma(\varepsilon_a)$}}}
\newcommand{\sigeo}{{\mbox{$\Sigma(\varepsilon)$}}}
\newcommand{\dsige}{{\mbox{$\Sigma'(\varepsilon_a)$}}}
\newcommand{\ddsige}{{\mbox{$\Sigma''(\varepsilon_a)$}}}

\newcommand{\dsigeo}{{\mbox{$\Sigma'(\varepsilon)$}}}
\newcommand{\ddsigeo}{{\mbox{$\Sigma''(\varepsilon)$}}}

\newcommand{\intr}{ \int \limits_{ - \infty}^{\infty} }

\newcommand{\vmel}[2]{\mel{#1}{\Umag}{#2}}

\newcommand{\mel}[3]{\langle #1|#2|#3\rangle}

\AtBeginDocument{%

}
\begin{document}

\title{QED effects in quadratic Zeeman splitting \\ in highly charged hydrogen-like ions}

\author{V.~A.~Agababaev}
\affiliation{School of Physics and Engineering, ITMO University, Kronverkskiy pr. 49, 197101 St. Petersburg, Russia}
\author{E.~A.~Prokhorchuk}
\affiliation{Department of Physics, St. Petersburg State University, Universitetskaya nab. 7/9, 199034 St. Petersburg, Russia}
\author{D.~A.~Glazov}
\affiliation{School of Physics and Engineering, ITMO University, Kronverkskiy pr. 49, 197101 St. Petersburg, Russia}
\author{A.~V.~Malyshev}
\affiliation{Department of Physics, St. Petersburg State University, Universitetskaya nab. 7/9, 199034 St. Petersburg, Russia}
\affiliation{Petersburg Nuclear Physics Institute named by B. P. Konstantinov of
National Research Centre ``Kurchatov Institute'', Gatchina, Russia}
\author{V.~M.~Shabaev}
\affiliation{Department of Physics, St. Petersburg State University, Universitetskaya nab. 7/9, 199034 St. Petersburg, Russia}
\affiliation{Petersburg Nuclear Physics Institute named by B. P. Konstantinov of
National Research Centre ``Kurchatov Institute'', Gatchina, Russia}
\author{A.~V.~Volotka}
\affiliation{School of Physics and Engineering, ITMO University, Kronverkskiy pr. 49, 197101 St. Petersburg, Russia}

\begin{abstract}
We present {\textit{ab initio}} calculations of one-electron quantum electrodynamical corrections to the second-order Zeeman splitting for the $1s_{1/2}$, $2s_{1/2}$, and $2p_{1/2}$ states in highly charged hydrogen-like ions. The self-energy correction is evaluated using the rigorous QED approach. The vacuum-polarization correction is evaluated within the electric-loop approximation. Calculations are performed for the wide range of nuclear charge number: $Z = 14$--$92$.
\end{abstract}

%
%
\maketitle
\section{Introduction}
Over the last two decades, research on the Zeeman effect in highly charged ions has developed intensively. The experimental precision has reached the level of $10^{-10}$--$10^{-11}$ for the \g factor of hydrogen-, lithium-, and boron-like ions \cite{haeffner:00:prl,verdu:04:prl,sturm:11:prl,sturm:13:pra,wagner:13:prl,Arapoglou:2019:PRL,glazov:19:prl,Heisse:2023:PRL}, including hydrogen- and boron-like tin \cite{Morgner:23:nature,Morgner:2025:PRL}, the heaviest to date. One of the most prominent results achieved by the community based on these studies is the most accurate determination of the electron mass from the $g$ factor of light hydrogen-like ions \cite{sturm:14:n,CODATA:2025:025002}.
The $g$-factor measurements in middle-$Z$ hydrogen- and lithium-like ions have provided the most stringent test of bound-state quantum electrodynamical (QED) theory in the presence of magnetic field \cite{wagner:13:prl,volotka:14:prl,koehler:16:nc,yerokhin:17:pra,Morgner:23:nature}. The relativistic nuclear recoil effect in highly charged ions represents the bound-state QED beyond the Furry picture in the strong-coupling regime \cite{malyshev:17:jetpl,shabaev:17:prl,Shabaev:18:pra}, and has recently been accessed in the isotopic-shift measurements \cite{koehler:16:nc,Sailer:2022:nature}.
Anticipated investigations on few-electron ions can serve for the determination of the fine-structure constant $\alpha$ \cite{shabaev:06:prl,volotka:14:prl-np,yerokhin:16:prl}, as well as for probing effects beyond the Standard Model \cite{Debierre:2020:PLB,Debierre:2022:PRA,Sailer:2022:nature,Akulov:25:a}.

Studies of the quadratic Zeeman effect have started more than 80 years ago from the discovery by Jenkins and Segr\'e \cite{jenkins:pr:1939} and a quantum-mechanical explanation by Shiff and Snyder \cite{shiff:pr:1939}. 
Numerous studies of this effect have been carried out in atoms, molecules, and exotic systems, such as positronium \cite{Garton:1969,Feinberg:pra:1990,Raoult:2005, Numazaki:pra:2010, Ivanov:2022:PRE}.  
Second-order Zeeman effect turns out to be prominent in solids and Bose-Einstein condensates \cite{Litvinenko:2016:SST,Veissier:2016:PRB, Davidson:2023:PRB,Qu:2020:PRL,Cheng-jie:2022:RP}. 
Zeeman splitting also plays an important role in astrophysics for determination of the magnetic field strength \cite{Babcock:1967:P,Crutcher:2019:FASS,Huerta:2024:RSI}. Its non-linearity is especially important in observation of such strong-field objects as magnetic white dwarfs (up to $10^{5}$ T) and magnetars (up to $10^{11}$ T). 

An important role of the second-order Zeeman effect is also recognized in atomic-clock engineering. 
To eliminate the linear effect in atoms and ions, transitions between levels with the total-angular-momentum projection $M_F = 0$ are usually chosen. However, the quadratic effect is still present, and its theoretical description is in demand \cite{Lu:2022:CPS, Steinel:2023:PRL, Qiao:2024:PRX}. The quadratic Zeeman shift has been directly measured for clock transitions in ${}^{27}\mathrm{Al}^+$ \cite{Brewer:2019:PRA},  ${}^{87}\mathrm{Rb}$  \cite{Wu:2013:JPB},  ${}^{87}\mathrm{Sr}$ \cite{Aeppli:2024:PRL,Lu:2020:AS,Bothwell:2019:metrologia}, and $^{171}{\mathrm{Yb}}^{+}$ \cite{Godun:2014:PRL}. 
Special attention is also paid to the quadratic Zeeman effect when developing atomic interferometers \cite{Hu:2017:PRA, Bao:2018:PRL, Ji:2024:PRA}.

In highly charged ions, the non-linear effects in Zeeman splitting are enhanced for fine-structure levels, such as $2p_j$ states in boron-like ions. Recently, the $g$ factors of both ground $2p_{1/2}$ and excited $2p_{3/2}$ states in boron-like argon ${}^{40}\mathrm{Ar}^{13+}$ have been determined with high precision in a single-ion Penning trap experiment \cite{Arapoglou:2019:PRL,Egl:2019:PRL}. These measurements were sensitive to the second- and third-order contributions in the magnetic field predicted in \cite{lindenfels:13:pra,glazov:13:ps,varentsova:18:pra}. An independent quantum-logic laser-spectroscopy experiment provided the ratio of these $g$ factors \cite{Micke:2020:Nature}. Lately, this method has been applied to measure the $g$ factor of the excited $^3P_1$ state and the corresponding second-order effect in carbon-like calcium $\mathrm{Ca}^{14+}$ \cite{spiess:25}. Prospects for future Zeeman and hyperfine splitting measurements within laser-microwave double-resonance spectroscopy have also been discussed \cite{lindenfels:13:pra,vogel:18:ap}. 

Theoretical investigations of the $g$ factor of boron-like ions have been accomplished in Refs.~\cite{lindenfels:13:pra,glazov:13:ps,verdebout:14:adndt,shchepetnov:15:jpcs,marques:16:pra,agababaev:18:jpcs,Agababaev:19:XRS,Maison:19:pra,cakir:20:pra}. Due to neglect of the negative spectrum, the results of MCDF calculations \cite{verdebout:14:adndt,marques:16:pra} are incomplete, as became clear from comparison with the experimental results \cite{Arapoglou:2019:PRL,Egl:2019:PRL,Micke:2020:Nature,Morgner:2025:PRL}.
Second- and third-order contributions to the Zeeman splitting in boron-like ions were considered in Refs.~\cite{lindenfels:13:pra,glazov:13:ps,Agababaev:17:NIMB,Varentsova:17:NIMB,varentsova:18:pra}. We also mention our recent study of the Zeeman splitting of the excited $2p_j$ states in lithium-like ions, including the non-linear effects \cite{zinenko:25}. These studies focus mainly on the interelectronic interaction, which is significant for non-linear contributions because of its strong impact on the fine-structure transition energy. Meanwhile, the effects of quantum electrodynamics (QED) are also very important in general for highly charged ions.

In the present work, we focus on the rigorous treatment of the first-order QED corrections to the quadratic Zeeman effect. To the best of our knowledge, the \textit{ab initio} fully relativistic consideration of the self-energy correction to this effect has never been done previously, except for Ref.~\cite{Agababaev:17:NIMB}, where the incorrect result was presented due to a mistake in the calculations. We present the formal expressions for the self-energy correction, discuss the cancellation of the ultraviolet and infrared divergences, and perform pilot calculations for the $1s$, $2s$, and $2p_{1/2}$ states in the wide range of highly charged hydrogen-like ions. The vacuum-polarization correction is also considered, within the electric-loop approximation only. Since the contributions of the magnetic loop with one or two magnetic-field interactions are presumably small, their consideration is beyond the scope of the present study. The calculations of the QED corrections to the quadratic Zeeman effect in lithium- and boron-like ions taking into account the interelectronic interaction will be carried out in the subsequent works.

Throughout the paper, we use relativistic units $\hbar=1$, $c=1$, $m_e=1$, and the Heaviside charge unit $e^2=4\pi\alpha$, where $e<0$ is the electron charge, $\mub = |e|/2$ is the Bohr magneton. 

\section{Quadratic Zeeman effect\label{sec:theory}}

Our physical model is a highly charged hydrogen-like ion with an infinitely heavy spinless nucleus placed in the external homogeneous magnetic field directed along the \textit{z}-axis, ${\bm B} = B {\bm e}_z$. All the expressions for the QED contributions presented below are applicable to any state of hydrogen-like ions and can also be used straightforwardly to treat the QED corrections to the quadratic Zeeman effect in ions with one electron over the closed shells, e.g., for lithium- and boron-like ions. For demonstration purposes, in the present work, we consider the ground $1s_{1/2}$ state as well as the excited $2s_{1/2}$ and $2p_{1/2}$ states.  

Assuming that the magnetic field is relatively weak, the energy of the one-electron state $|a\rangle$ can be expanded in a series,
\begin{equation}
\label{eq:pt1}
    E = E^{(0)} + E^{(1)} + E^{(2)} + \dots 
\,,
\end{equation}
where the numbers in parentheses correspond to the power of $B$. In its turn, each term in Eq. (\ref{eq:pt1}) can be expanded in {the fine-structure constant $\alpha$} as:
\begin{equation}
\label{eq:expanded}
    { E^{(m)} = \sum_{k} E^{(m|k)} \,. }
\end{equation}
 In case of many-electron systems, the formula (\ref{eq:expanded}) would acquire an additional index to take into account an expansion in powers of the interelectronic interaction. Within the rigorous QED approach employed in the present work, the zeroth-order approximation, {$E^{(0|0)}$}, is given by the eigenvalue $\varepsilon_a$ of the Dirac Hamiltonian, which includes the Coulomb potential of the nucleus $\Vnuc(r)$:
\begin{equation}
\label{eq:dirac}
    h_{\rm D}|a\rangle \equiv [\balpha \cdot{\bm p} + \beta + \Vnuc(r)]|a\rangle = \varepsilon_a|a\rangle 
\,,
\end{equation}
where $\balpha$ and $\beta$ are the Dirac matrices and ${\bm p}=-i\bnabla$ is the momentum operator. Perturbation theory in $\alpha$ can be conveniently constructed within the
two-time Green's-function method \cite{Shabaev:02:pr}. The terms {$E^{(0|1)}$ and $E^{(0|2)}$} correspond to the one- and two-loop QED contributions, respectively \cite{Mohr:1974:AnDP,Soff:1988:PRA,Manakov:1989:JETP,Jentshura:1999:PRL,Jentshura:2001:PRA,Sapirstein:2003:PRA,Yerokhin:2003:EPJD,Yerokhin:2024:PRL,Yerokhin:2025:042820}. 

The interaction with the magnetic field ${\bm B}$ reads as $V_{\rm magn}=-e\, \balpha\cdot {\bm A}_{\rm cl}$, where the classical vector potential for the homogeneous field is given by ${\bm A}_{\rm cl}=[{\bm B}\times{\bm r}]/2$, ${\rm rot}{\bm A}_{\rm cl} = {\bm B}$.  
In Eq.~(\ref{eq:pt1}), the contribution of  first order in the magnetic field, $E^{(1)}$, can be conveniently expressed in terms of the dimensionless factor $g^{(1)}$ by extracting the product of the Bohr magneton $\mub$ and the magnetic induction $B$.
The factor $g^{(1)}$ is proportional to the projection $M_J$ of the total angular momentum $J$, it is related to the {conventionally defined $g$ factor} as $g^{(1)} = g M_J${, where $g$ is independent on $M_J$}. The expansion in Eq.~(\ref{eq:expanded}) for energies naturally leads to a similar series for $g^{(1)}$,
\begin{equation}
\label{eq:gfactor}
    { E^{(1)} =\mub B g^{(1)}(M_J) = \mub B \sum_{k}  g^{(1|k)}(M_J) \,. }
\end{equation}
The term {$g^{(1|0)}$} is just the Dirac \g factor times $M_J$. The {contribution $g^{(1|1)}$ is} related to the corresponding one-loop QED  corrections {\cite{Yerokhin:2002:PRL, Karshenboim:2001:JETP}}. The term {$g^{(1|2)}$} refers to the two-loop QED correction, for the latest progress on which see Refs.~\cite{Yerokhin:13:PRA, Debierre:21:PRA, Sikora:2025:123001}.

The next term in Eq.~(\ref{eq:pt1}), $E^{(2)}$, corresponds to the quadratic Zeeman effect. Similar to $E^{(1)}$, we express it by the dimensionless factor $g^{(2)}$:
\begin{equation}
\label{eq:d2_def}
    { E^{(2)} =\frac{(\mub B)^2 }{m_e c^2} g^{(2)}(M_J)  =\frac{(\mub B)^2}{m_e c^2} \sum_{k}  g^{(2|k)}(M_J) \,, }
\end{equation}
where the factor $m_e c^2$ (equal to 1 in the units employed) is written explicitly for clarity.
The dependence of $\gbb$ on the projection $M_J$ is more complicated than that of $g^{(1)}$, so in general case it can not be factorized. Meanwhile, it depends only on the absolute value of $M_J$. Therefore, for the states with $J=1/2$ considered in the present work, the quadratic effect does not influence the Zeeman splitting. Still, it slightly shifts the fine-structure transition energy in boron-like ions~\cite{Egl:2019:PRL}. The case of $J>1/2$, e.g., the $2p_{3/2}$ state{, where the quadratic effect for $|M_J|=1/2$ and $|M_J|=3/2$ has to be evaluated separately}, will be considered elsewhere.

The leading-order contribution to $\gbb$ is given by:
\begin{equation}
\label{eq:gbb-dirac}
    { \gqd{0} = \sumnn\frac{\mel{a}{\Umag}{n}\mel{n}{\Umag}{a}}{\varepsilon_a - \varepsilon_{n}} \,, }
\end{equation}
where the summation runs over the complete Dirac spectrum excluding the reference state $\ket{a}$, and we have introduced the dimensionless operator $\Umag = \Vm /(\mub B) = [\bm{r}\times \balpha]_z$. {Here and in what follows, we omit the dependence of $g^{(2)}$ on $M_J$ for brevity.} The correction of first order in $\alpha$, {$\gqd{1}$}, is given by the sum of the self-energy and vacuum-polarization corrections,
\begin{equation}
    { \gqd{1} \equiv \dgbbse + \dgbbvp \,, }
\end{equation}
and represents the main topic of the present work. In the following section, we proceed with the discussion of the corresponding contributions.

{As an alternative to the perturbation theory in the magnetic field outlined above, one can consider an all-order approach in $V_{\rm magn}$, also termed as a finite-field method. In this case, instead of Eq.~(\ref{eq:dirac}),  we have the Dirac equation for a particle placed in the combined Coulomb nuclear and external magnetic fields as the starting point:
\begin{align}
\label{eq:dirac_2}
\tilde{h}_{\rm D}|\tilde{a}\rangle \equiv [\balpha \cdot{\bm p} + \beta + \Vnuc(r) 
+ V_{\rm magn}({\bm r})] |\tilde{a}\rangle = \tilde{\varepsilon}_a|\tilde{a}\rangle 
\,,
\end{align}
where the tilde sign is introduced to distinguish Hamiltonian, wave functions, and energies obtained in the presence of the magnetic field from their counterparts obtained in its absence. In this formalism, the factor factor $g^{(2)}$ reads as
\begin{align}
\label{eq:g2_2}
g^{(2)} = \frac{1}{2} \frac{m_e c^2}{\mub^2} \left. \frac{\partial^2}{\partial B^2} \right|_{B=0} E \, ,
\end{align}
where $E$ is the total energy of the ion in Eq.~(\ref{eq:pt1}) and the magnetic field $B$ should be set to zero after taking the partial derivatives. In particular, the leading-order contribution considered in Eq.~(\ref{eq:gbb-dirac}) can be obtained from Eq.~(\ref{eq:g2_2}) by replacing $E$ with $\tilde{\varepsilon}_a$.
}

{We should also note that, to some extent, one can go beyond the external-field approximation corresponding to the infinitely heavy nucleus and derive corrections to certain contributions $E^{(m|k)}$ in Eq.~(\ref{eq:expanded}) in the form of a series in the electron-to-nucleus mass ratio~$m_e/M$. For instance, the fully relativistic description of the nuclear-recoil effect on energy levels in the absence of the magnetic field was given in Refs.~\cite{Shabaev:1985:TMP,Shabaev:1988:SJNP,shabaev:98:pra,Pachucki:1995:PRA,Adkins:2007:PRA}. The radiative correction to the nuclear-recoil contribution has been considered recently in Ref.~\cite{Pachucki:2024:PRA}. The QED theory of the nuclear-recoil effect on the $g$ factor was developed in Ref.~\cite{shabaev:01:pra}. The case of the quadratic Zeeman effect, however, is much more complicated~\cite{Pachucki:2008:012504}.
}

\subsection{Self-energy correction}
\hspace{\parindent}

{In the present work, in order to derive the formulas for the QED corrections to the quadratic Zeeman effect, we use the two-time Green's function method \cite{Shabaev:02:pr} and treat the magnetic field perturbatively. However, ideologically simpler formal expressions can be readily obtained in the framework of the alternative approach discussed at the end of the previous section. For instance, using Eq.~(\ref{eq:g2_2}), one easily obtains for the self-energy correction to the quadratic Zeeman effect
\begin{align}
\label{eq:g2_se:nonpert}
g^{\rm(2|SE)} = \frac{1}{2} \frac{m_e c^2}{\mub^2} \left. \frac{\partial^2}{\partial B^2} \right|_{B=0} \tilde{E}^{\rm (SE)} \, ,
\end{align}
where
\begin{align}
\label{eq:se:nonpert_1}
\tilde{E}^{\rm (SE)} &= \langle \tilde{a} | \tilde{\Sigma}_R(\tilde{\varepsilon}_a)| \tilde{a} \rangle \, 
\end{align}
is the self-energy correction to the energy level of the ion placed in the external magnetic field. The one-loop self-energy operator in the magnetic field in Eq.~(\ref{eq:se:nonpert_1}) reads as
\begin{align}
\label{eq:se:nonpert_2}
\tilde{\Sigma}_R(\varepsilon,{\bm r}_1,{\bm r}_2) &=
2i\alpha \int \! d\omega \, D^{\mu\nu}(\omega,{\bm r}_{12}) \alpha_\mu 
\tilde{G}(\varepsilon-\omega,{\bm r}_1,{\bm r}_2)\alpha_\nu - \beta \delta m 
\end{align}
where ${\bm r}_{12}={\bm r}_{1}-{\bm r}_{2}$, $\alpha^\mu=(1,\balpha)$, $D_{\mu\nu} (\omega,{\bm r}_{12})$ is the photon propagator, $\delta m$ is the mass counterterm, and $\tilde{G}=(\varepsilon - \tilde{h}_{\rm D})^{-1}$ is the Green's function for the electron in the combined nuclear and magnetic fields.
The formula (\ref{eq:g2_se:nonpert}) is difficult to apply in calculations. In principle, one can replace the Green's function, wave functions, and energy with their perturbative series up to the second order in the magnetic field, and rewrite this expression in a form suitable for numerical evaluation. However, we prefer to use the two-time Green's function method instead to obtain the same result.
}

\begin{figure}[ht]
\begin{center}
\hspace{-1cm}
\includegraphics[width=0.85\linewidth]{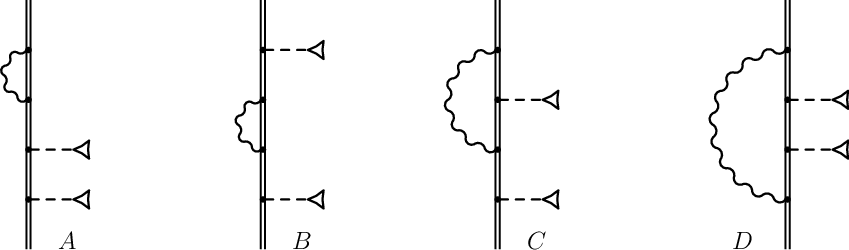}\\
\caption{Self-energy diagrams. The double lines correspond to the electron propagators in the nuclear potential~$V_{\rm nuc}$. The wavy lines denote the photon propagators. The mass-counterterm diagrams are omitted.}
\label{fig:se}
\end{center}
\end{figure}

{Within the perturbative-in-magnetic-field approach, t}he self-energy correction to the quadratic Zeeman effect is given by the gauge-invariant set of the four topologically non-equivalent diagrams shown in Fig.~\ref{fig:se}. Similar diagrams with one of the vertices replaced by the hyperfine-interaction operator describe the self-energy correction to the nuclear-magnetic shielding \cite{yerokhin:11:prl,yerokhin:2012:pra}. Self-energy diagrams with two vertices were also considered for the parity non-conserving amplitude in heavy alkali-metal atoms \cite{shabaev:05:pra}. Two-electron self-energy diagrams for the $g$ factor and hyperfine splitting \cite{volotka:09:prl,Glazov:10:pra,yerokhin:20:pra,glazov:19:prl,kosheleva:22:prl} are very close in structure as well. 
The relevant mass-counterterm diagrams are not shown explicitly in Fig.~\ref{fig:se}, but they are used in the renormalization procedure.
The total self-energy correction to $g^{(2)}$ is given by the sum of ten terms,
\begin{equation}
\label{eq:se:1}
    \dgbbse = \sum_{X \in \Omega} \dgqse{X}
\,,
\end{equation}
which we enumerate with the index $X$ running over the set $\Omega=\{A,B,C,D,G1,G2,H1,H2,I1,I2\}$.
The so-called irreducible contributions of the diagrams A, B, C, and D, where the intermediate states differ from the reference state $|a\rangle$, are labeled with the same letter. The corresponding expressions are 
\begin{equation}
\label{eq:se:start}
\dgqse{A}=2\,\summn \frac {\mel{a}{\sige}{n_1} \vmel{n_1}{n_2}\vmel{n_2}{a}}{\depsn{1}\depsn{2}},
\end{equation}
\begin{equation}
\dgqse{B}=\summn \frac{\vmel{a}{n_1} \mel{n_1}{\sige}{n_2} \vmel{n_2}{a}}{\depsn{1}\depsn{2}},
\end{equation}
\begin{equation}
\label{eq:C}
\dgqse{C}=2 \sumnn\frac{\mel{a}{\gum}{n}\vmel{n}{a}}{\varepsilon_a-\varepsilon_n},
\end{equation}
\begin{equation}
\label{eq:se:start_D}
\dgqse{D} = \mel{a}{\lv}{a}.
\end{equation}
Here and below, a prime on the sum indicates that the terms with vanishing denominators are excluded from the summation over the complete Dirac spectrum.
In addition, the following notations are employed in Eqs.~(\ref{eq:se:start})--(\ref{eq:se:start_D}),
\begin{equation}
\label{eq:oper-beg}
{ \langle a|\sigeo{}|b\rangle\equiv\frac{i}{2\pi}\intr d\omega\sum_n\frac{\langle an|I(\omega)|nb\rangle}{\varepsilon-\omega-\varepsilon_n^-}
}
\end{equation}
for the self-energy matrix element,
\begin{equation}
\label{eq:gumo}
{\langle a|\gumo|b\rangle\equiv\frac{i}{2\pi}\intr d\omega\sum_{n_1,n_2}\frac{\langle an_2|I(\omega)|n_1b\rangle\langle n_1|\Umag|n_2\rangle}{(\varepsilon-\omega-\varepsilon_{n_1}^-)(\varepsilon-\omega-\varepsilon_{n_2}^-)}}
\end{equation}
for the single-vertex matrix element, and
\begin{equation}
\label{eq:oper-end}
\langle a|\Lambda[\Umag,\Umag](\varepsilon)|b\rangle \equiv \frac{i}{2\pi}\intr d\omega\sum_{n_1,n_2,n_3}\frac{\langle a n_3 |I(\omega)|n_1 b\rangle\langle n_1|\Umag|n_2\rangle \langle n_2|\Umag|n_3\rangle}{(\varepsilon-\omega-\varepsilon_{n_1}^-)(\varepsilon-\omega-\varepsilon_{n_2}^-)(\varepsilon - \omega - \varepsilon_{n_3}^-)}
\end{equation}
for double-vertex matrix element. In Eqs.~(\ref{eq:oper-beg})--(\ref{eq:oper-end}) and what follows, $\varepsilon_{n}^- \equiv \varepsilon_{n}(1-i0)$ provides the proper bypass of all the singularities in the complex $\omega$ plane, and the operator $I(\omega)$ is defined by
\begin{align}
\label{eq:I_omega}
    I(\omega,{\bm r}_{12}) = e^2 \alpha^\mu \alpha^\nu D_{\mu\nu} (\omega,{\bm r}_{12}) 
\,.
\end{align}

The remaining parts of the self-energy correction, so-called reducible contributions, are written as follows,
\begin{equation}
\label{eq:se:G1}
\dgqse{G1}{} = -2 \vmel{a}{a}\sumnn  \frac{\mel{a}{\sige}{n} \vmel{n}{a}}{\depsn{}^2},
\end{equation}
\begin{equation}
\dgqse{G2}{} = - \mel{a}{\sige}{a} \sumnn \frac{\vmel{a}{n} \vmel{n}{a}}{\depsn{}^2},
\end{equation}
\begin{equation}
\dgqse{H1}{} = 2 \vmel{a}{a} \sumnn  \frac{\mel{a}{\dsige}{n} \vmel{n}{a}}{\varepsilon_a-\varepsilon_n},
\end{equation}
\begin{equation}
\dgqse{H2}{} = \mel{a}{\dsige}{a} \sumnn  \frac{\vmel{a}{n} \vmel{n}{a}}{\varepsilon_a-\varepsilon_n},
\end{equation}
\begin{equation}
\dgqse{I1}{} = \mel{a}{\dgu}{a}\vmel{a}{a},
\end{equation}
\begin{equation}
\label{eq:se:end}
\dgqse{I2}{} = \frac{1}{2} \mel{a}{\ddsige}{a}\vmel{a}{a}\vmel{a}{a}.
\end{equation}
Here we use the following notations for the derivatives of the operators (\ref{eq:oper-beg})--(\ref{eq:oper-end}),
\begin{equation}
    \langle a|\dsigeo{}|b\rangle\equiv-\frac{i}{2\pi}\intr d\omega\sum_n\frac{\langle an|I(\omega)|nb\rangle}{(\varepsilon-\omega-\varepsilon_n^-)^2}
\,,
\end{equation}
\begin{equation}
    \langle a|\ddsigeo|b\rangle\equiv\frac{i}{\pi}\intr d\omega\sum_n\frac{\langle an|I(\omega)|nb\rangle}{(\varepsilon-\omega-\varepsilon_n^-)^3}
\,,
\end{equation}
\begin{align}
    \langle a|\Gamma^{\prime}[\Um](\varepsilon)|b\rangle \equiv -\frac{i}{2\pi}\intr d\omega\sum_{n_1,n_2}\Bigg[&\frac{\langle an_2|I(\omega)|n_1 b\rangle\vmel{n_1}{n_2}}{{(\varepsilon-\omega-\varepsilon_{n_1}^-)}^2(\varepsilon - \omega - \varepsilon_{n_2}^-)}+
\nonumber\\
    + & \frac{\langle an_2|I(\omega)|n_1 b\rangle\vmel{n_1}{n_2}}{{(\varepsilon-\omega-\varepsilon_{n_1}^-)}(\varepsilon - \omega - \varepsilon_{n_2}^-)^2}\Bigg]\frac{}{}
\,.
\end{align}

All the contributions except for $D$, $I1$, and $I2$ contain ultraviolet divergences. Regularization of these divergences is discussed in Appendix~\ref{sec:app:1}.
The terms $\dgqse{D}{}$, $\dgqse{H2}{}$, $\dgqse{I1}{}$, and $\dgqse{I2}{}$ also possess infrared divergences, they cancel out in the sum. 
The problem of treating the infrared divergences is considered in Appendix~\ref{sec:app:irr}. 

\subsection{Vacuum-polarization correction}
\begin{figure}[ht]
\begin{center}
\includegraphics[width=0.8\linewidth]{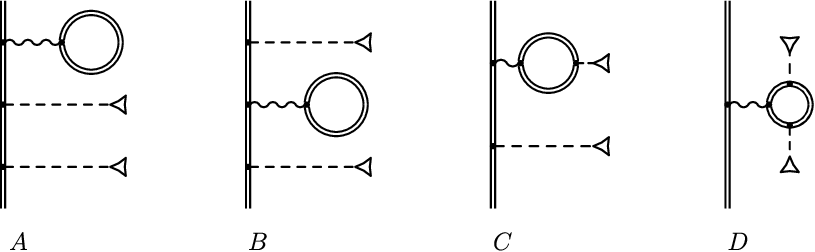}\\
\caption{Vacuum-polarization diagrams. The notations are the same as in Fig.~\ref{fig:se}.}
\label{fig:vp}
\end{center}
\end{figure}

The diagrams describing the vacuum-polarization correction to the quadratic Zeeman effect are shown in Fig.~\ref{fig:vp}. Drawing the formal analogy with the self-energy diagrams in Fig.~\ref{fig:se}, one can express the total vacuum-polarization contribution as follows,
\begin{equation}
\label{eq:vp_x}
    \dgbbvp = \sum_{X \in \Omega} \dgqvp{X}.
\end{equation}
However, since the fermion loops in the diagrams in Fig.~\ref{fig:vp} do not depend on energy parameter, the derivatives over this parameter yield zero. Therefore, the terms $\dgqvp{X}$ with $X$ in $\{H1,H2,I1,I2\}$ vanish identically.
The electric-loop diagrams, where the magnetic interaction does not enter the fermion loop, give the dominant contribution to the vacuum-polarization correction to the \g factor of hydrogen- and lithium-like ions \cite{persson:97:PRA, Karshenboim:2001:JETP,glazov:04:pra}. We suppose that the same holds true also for $\gbb$. For this reason, in the present work, we limit the vacuum-polarization correction by the electric-loop diagrams $A$ and $B$, and omit the contribution of the diagrams $C$ and $D$. So, the index $X$ in Eq.~(\ref{eq:vp_x}) runs over $\Omega' = \{A, B, G1, G2\}$.
To derive the formal expressions for the corresponding diagrams, we use the two-time Green's function method \cite{Shabaev:02:pr}.
The irreducible contributions are given by:
\begin{equation}
\label{eq:vp1}
    \dgqvp{A}=2\,\summn \frac {\langle a|U_{\mathrm{VP}}|n_1\rangle \langle n_1 |\Umag| n_2\rangle \langle n_2|\Umag|a\rangle}{(\varepsilon_a-\varepsilon_{n_1})(\varepsilon_a-\varepsilon_{n_2})},
\end{equation}
\begin{equation}
\label{eq:vp2}
    \dgqvp{B}=\summn \frac{\langle a|\Umag|n_1\rangle \langle n_1|U_{\mathrm{VP}}|n_2 \rangle \langle n_2|\Umag|a \rangle}{(\varepsilon_a - \varepsilon_{n_1})(\varepsilon_a - \varepsilon_{n_2})},
\end{equation}
and the corresponding reducible contributions are:
\begin{equation}
\dgqvp{G1}= - 2\,\langle a|\Umag|a \rangle \sumnn \frac {\langle a|U_{\mathrm{VP}}|n\rangle \langle n|\Umag| a\rangle}{(\varepsilon_a-\varepsilon_{n})^2},
\end{equation}
\begin{equation}
\label{eq:vp4}
\dgqvp{G2}= - \langle a|U_{\mathrm{VP}}|a \rangle \sumnn \frac{\langle a|\Umag|n \rangle  \langle n|\Umag|a \rangle}{(\varepsilon_a - \varepsilon_{n})^2}.
\end{equation}
In Eqs.~(\ref{eq:vp1})--(\ref{eq:vp4}), $U_{\mathrm{VP}}$ is the vacuum-polarization potential. The leading-order in $\alpha Z$ contribution to $U_{\mathrm{VP}}$ is given by the Uehling potential \cite{Uehling:1935:55,Serber:1935:49},
\begin{align}
\label{eq:Ue}
    U_{\mathrm{VP}}^{\mathrm{Ue}} (r) = -\frac{2\alpha^2 Z}{3r}
        \int\limits_0^\infty \! dr' \, r' \rho(r') \left[ K_0(2|r-r'|) -K_0(2|r+r'|) \right] 
\,,
\end{align}
where
\begin{align}
    K_0(x) = \int\limits_1^\infty \! dt \, e^{-xt} \left(\frac{1}{t^3}+\frac{1}{2t^5}\right) \sqrt{t^2-1} 
\,, 
\end{align}
and the nuclear charge density $\rho$ is normalized by the condition $\int \!d{\bm r}\, \rho  = 1$. The Uehling potential (\ref{eq:Ue}) can easily be evaluated employing the approximate formulas from Ref.~\cite{Fullerton:1976:1283}. The calculation of the higher-order in $\alpha Z$ contribution to $U_{\mathrm{VP}}$, the Wichmann-Kroll potential~$U_{\mathrm{VP}}^{\mathrm{WK}}$, is a more complicated problem~\cite{Soff:1988:PRA,Manakov:1989:JETP,Persson:1993:2772}, see also Refs.~\cite{Artemyev:1997:3529, Salman:2023:012808, Ivanov:2024:032815, Ivanov:25:pra}. In the present work, we take it into account using the approximate formulas derived for a point nucleus in Ref.~\cite{Fainshtein:1991:559}.
To complete the discussion of the vacuum-polarization corrections, we note that the magnetic-loop contribution $\dgqvp{C}$ vanishes identically within the Uehling (free-loop) approximation, see, e.g., Ref.~\cite{Beier:2000:PR}.


\section{Results and discussion}

In this section, we present the results of our calculations of the quadratic Zeeman effect on the $1s$, $2s$, and $2p_{1/2}$ states of hydrogen-like ions in the wide range of nuclear charge numbers from $Z=14$ to $Z=92$. All calculations are carried out for extended nuclei, the Fermi model is employed to describe the nuclear-charge distribution. 

Before proceeding to the discussion of the QED corrections, self-energy~$\dgqse{}$ and vacuum polarization~$\dgqvp{}$, which are the main goal of the present work, let us focus on the leading contribution, {$\gqd{0}$}, given by Eq.~(\ref{eq:gbb-dirac}). The summation over the Dirac spectrum in Eq.~(\ref{eq:gbb-dirac}) 
is performed using finite-basis sets constructed from B splines~\cite{Johnson:1988:307,Sapirstein:1996:5213} within the dual-kinetic-balance approach \cite{shabaev:04:prl}. The results for {$\gqd{0}$} are shown in Table~\ref{tab:gbb}. It can be seen that for the $2p_{1/2}$ state the quadratic Zeeman effect is more pronounced than for the $s$ states. The reason for this is the proximity of the $2p_{1/2}$ and $2p_{3/2}$ levels which leads to a small denominator in Eq.~(\ref{eq:gbb-dirac}). As a result, the behavior of the leading contribution {$\gqd{0}$} for the $2p_{1/2}$ state is determined by the fine-structure $2p_{1/2}-2p_{3/2}$ splitting. It scales approximately as $1/(\alpha Z)^4$, while for the $s$ states one obtains only a $1/(\alpha Z)^2$ scaling. We note that the smallness of energy denominators in the case of the $2p_{1/2}$ state also manifests itself when calculating the QED corrections. 

The formal expressions for the self-energy correction to the quadratic Zeeman effect are given by Eqs.~(\ref{eq:se:start})--(\ref{eq:se:start_D}) and (\ref{eq:se:G1})--(\ref{eq:se:end}). In Appendix~\ref{sec:app:1}, it is shown that the total self-energy correction is ultraviolet finite. According to the well-established renormalization procedures \cite{yerokhin:99:pra,Yerokhin:2004:052503,yerokhin:10:pra}, 
the bound-electron Green’s functions, entering the self-energy loops, are expanded in terms of the binding potential $V_{\rm nuc}$, in order to separate out all ultraviolet-divergent contributions. The latter should be treated in momentum space, where the divergences can be covariantly regularized and explicitly canceled. The remaining part of the self-energy correction is evaluated in the coordinate space using the partial-wave expansion for the photon and electron propagators, and this is the most time-consuming part of the calculations.

An example of self-energy correction calculations for the $1s$, $2s$, and $2p_{1/2}$ states in hydrogen-like argon ($Z=18$) is shown in Tables~\ref{tab:details_1s}, \ref{tab:details_2s}, and \ref{tab:details_2p1}, respectively. The columns ``$A$'', ``$B$'', etc. correspond to the terms $\dgqse{A}{}$, $\dgqse{B}{}$, etc. defined by Eqs.~(\ref{eq:se:start})--(\ref{eq:se:start_D}) and (\ref{eq:se:G1})--(\ref{eq:se:end}), while the last column ``Sum'' shows the total self-energy contribution $\dgbbse$ in Eq.~(\ref{eq:se:1}). For the sake of convenience, the term $\dgqse{D}{}$ is placed last. In Tables~\ref{tab:details_1s}--\ref{tab:details_2p1}, the first rows labeled ``Free'' present the contributions evaluated in the momentum space after applying the renormalization procedure. Since the expressions $\dgqse{D}{}$, $\dgqse{I1}{}$, and $\dgqse{I2}{}$ are initially ultraviolet finite and therefore do not require renormalization, they do not contribute to these rows. The momentum-space contributions to the terms $\dgqse{A}{}$, $\dgqse{B}{}$, $\dgqse{G1}{}$, and $\dgqse{G2}{}$ are given by the generalization of the corresponding expressions presented, e.g., in Ref.~\cite{yerokhin:99:pra}. In the case of the terms $\dgqse{C}$, $\dgqse{H1}$, and $\dgqse{H2}$, we use the formulas similar to those given in Ref.~\cite{yerokhin:2012:pra}. 
The subsequent lines in the tables show the partial-wave expansion for the corresponding coordinate-space contributions. For each term $\dgqse{X}{}$, we add to the partial sums $S_k$ the individual terms with $|\kappa_{n_i}|<k$, $k=1,2,\ldots$, for all electron propagators inside the self-energy loops. The lines labeled $|\kappa|=k$ in Tables~\ref{tab:details_1s}--\ref{tab:details_2p1} give the increments of partial sums, i.e. $S_k - S_{k-1}$. For a better representation of the behavior of different terms with $|\kappa|$, we show extra digits for the individual values. As discussed in Section~\ref{sec:theory}, the terms $\dgqse{D}{}$, $\dgqse{H2}{}$, $\dgqse{I1}{}$, and $\dgqse{I2}{}$ taken separately are infrared divergent, although the total self-energy correction is not, see Appendix~\ref{ap:ir}. Since in the present work we focus on the states with a total angular momentum of $1/2$, the infrared divergences of the terms $\dgqse{H2}{}$, $\dgqse{I1}{}$, and $\dgqse{I2}{}$ manifest themselves only in the $|\kappa|=j+1/2=1$ contributions. The term $\dgqse{D}{}$ demonstrates the infrared behavior in two cases: (i) $\kappa_{n_1}=\kappa_{n_2}=\kappa_{n_3}\equiv\kappa_{a}$; (ii) $\kappa_{n_1}=\kappa_{n_3}\equiv\kappa_{a}$, while $\kappa_{n_2}=-\kappa_{a}+1=2$ for the $s$ states and $\kappa_{n_2}=-\kappa_{a}-1=-2$ for the $2p_{1/2}$ state. Therefore, case (i) corresponds to the $|\kappa|=1$ contributions, while case (ii) corresponds to the $|\kappa|=2$ contributions. We treat infrared divergences numerically by evaluating the divergent terms together. For this reason, the $|\kappa|=1$ contributions of the terms $\dgqse{D}{}$, $\dgqse{H2}{}$, $\dgqse{I1}{}$, and $\dgqse{I2}{}$ are omitted in Tables~\ref{tab:details_1s}--\ref{tab:details_2p1}, they are included in the $|\kappa|=2$ contributions of the term $\dgqse{D}{}$. We also note that, for the sake of better readability, we omit small contributions that should be represented by zeros in the significant digits shown in the tables.

The coordinate-space contributions to the self-energy correction are evaluated in the present work by means of two methods, which differ in the treatment of the electron propagators inside the self-energy loops. First, we use the finite-basis-set representation for the free- and bound-electron Green’s functions. Second, we find the Green's functions by solving a corresponding system of differential equations. Both methods are described, e.g., in Refs.~\cite{Yerokhin:2020:800}. The results, obtained by both approaches have been found in excellent agreement with each other. The summations over the electron spectra outside the loops are always performed using the finite-basis-set approach. The partial-wave expansion in the self-energy contributions is typically terminated at $|\kappa_{\rm max}|=24$. The remainders are obtained by polynomial (in $1/k$) least-squares fitting of the partial sums $S_k$ and our estimates for them are shown in the lines labeled ``$\sum_{|\kappa|\geqslant 25}$''. The extrapolation $|\kappa_{\rm max}|\to\infty$ is the main source of uncertainties shown in the parentheses. From Tables~\ref{tab:details_1s}--\ref{tab:details_2p1}, it can be seen that some terms converge with $|\kappa|$ rather rapidly, whereas the accurate evaluation of the others, e.g., the term~$\dgqse{D}{}$ for all states considered or the terms $\dgqse{B}{}$ and $\dgqse{G2}{}$ for the $2p_{1/2}$ state, is a challenging task. To improve the convergence of partial-wave expansion for these terms, we employ the generalization of the scheme proposed in Refs.~\cite{Artemyev:2007:173004, Artemyev:2013:032518} for the case of one-electron self-energy correction to energy levels. Namely, we revisit the expansion of the electron Green's functions in the self-energy loops in terms of $V_{\rm nuc}$ and, in addition to the ultraviolet-divergent contributions, subtract also the leading ultraviolet-finite contributions. For instance, for the term~$\dgqse{D}{}$ it implies a subtraction of a similar expression where all bound-electron propagators are replaced with their free-electron counterparts. After such subtractions, the resulting differences converge much better. The subtracted expressions are analyzed separately in the same coordinate space, but with their calculations extended up to $|\kappa_{\rm max}|=75$. Application of this approach allows us to significantly improve the precision of the obtained theoretical predictions. For all slowly-converging terms, the lines ``$\sum_{|\kappa|\geqslant 25}$'' show the values obtained by means of this method. In this case, the uncertainties shown include those due to the extrapolation of both the differences and the subtracted contributions. We note that the slow convergence of the total self-energy correction is entirely determined by the term $\dgqse{D}{}$, which involves two interactions with the magnetic field inside the loop. The discussed approach has been also applied, e.g., to the analysis of the individual terms $\dgqse{B}{}$ and $\dgqse{G2}{}$ in the case of the $2p_{1/2}$ state, but from Table~\ref{tab:details_2p1} one can conclude that their sum converges significantly better than these terms taken separately. In this context, numerous convergence-acceleration methods developed for different self-energy corrections should be mentioned \cite{Yerokhin:2005:042502, Sapirstein:2023:042804, Malyshev:2024:062802, Yerokhin:2024:PRL, Yerokhin:2025:012802}. In the future calculations, the similar schemes can be also applied to the self-energy correction to the quadratic Zeeman effect as well. Finally, the last lines in Tables~\ref{tab:details_1s}, \ref{tab:details_2s}, and \ref{tab:details_2p1} show the total self-energy corrections which are the sums of the momentum- and coordinate-space contributions. Note strong cancellation between these contributions, especially in the case of the $2s$ state.

In Table~\ref{tab:qed}, we present our results for the QED corrections to the quadratic Zeeman effect for the $1s$, $2s$, and $2p_{1/2}$ states of hydrogen-like ions in the range $14\leqslant Z \leqslant 92$. The calculations of the self-energy, SE, corrections are performed as demonstrated in Tables~\ref{tab:details_1s}, \ref{tab:details_2s}, and \ref{tab:details_2p1}. The uncertainties shown are purely numerical and are derived from the analysis of partial-wave-expansion convergence. The calculations of the vacuum-polarization, VP, corrections are carried out within the approximation discussed in Sec.~\ref{sec:theory}. Our treatment of the vacuum polarization is incomplete, but we expect the part accounted for in the present work to make the dominant contribution. From Table~\ref{tab:qed}, it can be seen that the vacuum-polarization correction is much smaller than the self-energy one. We do not give uncertainties for the vacuum-polarization corrections, since they can only be determined with confidence when the full consideration is completed. However, the presented values give a pretty clear understanding of the corrections under discussion, what meets the main goals of the present work. In Ref.~\cite{Agababaev:17:NIMB}, where the quadratic Zeeman effect in boron-like argon ($Z=18$) was studied, the QED corrections for the $2p_{1/2}$ state were evaluated for the Coulomb potential of the nucleus as well. The value of the vacuum-polarization correction, $0.2$, given there, is in agreement with the present value, $0.193$. In the case of the self-energy correction, due to some misprints in the numerical procedure a wrong value was obtained. In the present work, these inaccuracies have been fixed, the numerical procedures have been independently cross-checked, and the reliable and accurate data are obtained as a result.

As is well known, the QED corrections to the $g$ factor can be approximately treated by a set of effective operators, which can be derived by considering the interaction of the anomalous magnetic moment of electron with the magnetic and electric fields. 
The corresponding operators can be found in Refs. \cite{Hegstrom:73:pra, glazov:04:pra}.
In the case of hydrogen-like ions with spinless nuclei, only two operators contribute. The first operator, $H_1^{\rm rad}$, is linear in the external magnetic field, while the second operator, $H_2^{\rm rad}$, does not depend on ${\bm B}$, see Eqs.~(31) and (32) in Ref.~\cite{glazov:04:pra}, respectively. From the derivation, it follows that both operators are proportional to $g_{\rm free}-2$, where $g_{\rm free}$ is the free-electron $g$ factor. The complete description of the QED corrections to the bound-electron $g$ factor (to the leading orders in $\alpha Z$) can be obtained with both the first-order contribution in $H_1^{\rm rad}$ and the second-order cross contribution in $H_2^{\rm rad}$ and $V_{\rm magn}$~\cite{glazov:04:pra,glazov:19:prl}.

The natural question is whether these operators can be applied to the quadratic Zeeman effect calculations. For instance, in the recent work \cite{Gilles:2024:PRA} this effect was studied for various states in Ca$^{14+}$, Ni$^{12+}$, and Xe$^{q+}$ ions, and only the $H_1^{\rm rad}$ operator was taken into account, while the $H_2^{\rm rad}$ operator was omitted. We have undertaken the calculations for the $1s$, $2s$ and $2p_{1/2}$ states of hydrogen-like ions based on both these operators. Namely, the operators $H_1^{\rm rad}$ and $H_2^{\rm rad}$ have been considered together with $V_{\rm magn}$ within the perturbation theory to obtain the contributions of the second order in the magnetic field and of the first order in $g_{\rm free}-2$. 

The results of these calculations for the $2p_{1/2}$ state are presented in Table~\ref{tab:qed-nonrel}. The contributions arising from the operators $H_1^{\rm rad}$ and $H_2^{\rm rad}$ are shown separately in the second and third columns, respectively. These contributions partially cancel each other in the sum ``$H_1^{\rm rad}+H_2^{\rm rad}$'', which should be compared with the results of the rigorous QED calculations from Table~\ref{tab:qed}, also given here for convenience. The difference between the rigorous and approximate values, $\Delta$, and the ratio of $\Delta$ to the QED result are also shown. From Table~\ref{tab:qed-nonrel}, it can be seen that the QED corrections to the quadratic Zeeman effect for the $2p_{1/2}$ state obtained by applying the operators $H_1^{\rm rad}$ and $H_2^{\rm rad}$, {$g^{(2|1)}_{H_1^{\rm rad}+H_2^{\rm rad}}$}, are in reasonable agreement with the rigorously evaluated corrections {$g^{(2|1)}$}, and, as one would expect, the deviation between two methods tends to zero as $Z$ decreases. For the $s$ states, however, this is not the case. For example, for the $1s$ state in hydrogen-like silicon ($Z=14$), the contribution corresponding to the operator $H_1^{\rm rad}$ is $-0.001157$, while the contribution of the operator $H_2^{\rm rad}$ is much smaller and amounts to $-0.000005$. Moreover, the contribution of $H_1^{\rm rad}$ is almost independent of $Z$, while the contribution of $H_2^{\rm rad}$ grows in the absolute value, keeping the negative sign, and reaches $-0.000259$ for $Z=92$. A similar behavior is observed for the $2s$ state. We conclude that the approximate operators $H_1^{\rm rad}$ and $H_2^{\rm rad}$ cannot correctly describe the QED corrections to the quadratic Zeeman effect for the $s$ states. Detailed investigation of this issue is beyond the scope of the present paper and may be the subject of our future study. 

To complete the discussion of the latter issue, we mention that a special feature of the $2p_{1/2}$ state compared to the $s$ states is that the leading $\alpha Z$ behavior of $\gbb$ is determined by the fine-structure $2p_{1/2}-2p_{3/2}$ splitting. We have studied the nonrelativistic ($\alpha Z \rightarrow 0$) limit of the ratio {$g^{(2|1)}_{H_1^{\rm rad}+H_2^{\rm rad}} / g^{(2|0)}$} for the $2p_{1/2}$ state, and found that it is equal to $g_{\rm free}-2\approx \alpha/\pi$. This result motivates us to express the results for the QED corrections to the quadratic Zeeman effect on the $2p_{1/2}$ state in terms of the function $\faz $, defined as
\begin{equation}
\label{eq:faz}
    {  \gqd{1}_{2p_{1/2}}=\frac{\alpha}{\pi}{\gqd{0}} \faz \,. }
\end{equation}
{This is done in the last column of Table~\ref{tab:qed}. One can see that $\faz $ is indeed a slowly varying function of $Z$ which tends to 1 as $Z$ decreases.}

%
%

\section{Summary}

In the present work, the first-order one-electron QED corrections to the quadratic Zeeman splitting have been studied within the fully relativistic approach. The complete set of relevant self-energy Feynman diagrams and the dominant electric-loop vacuum-polarization part have been considered. For the formal self-energy expressions, accurate analysis of the ultraviolet and infrared divergences has been performed. Previously developed methods for self-energy and vacuum-polarization calculations to all orders in $\alpha Z$ have been extended and refined. Poor convergence of the partial-wave expansion has been tackled by accumulating the slowly converging tail within the free-electron approximation. The developed formalism has been applied to calculations of the quadratic Zeeman effect for the $1s$, $2s$, and $2p_{1/2}$ states in hydrogen-like ions in the wide range of nuclear charge numbers $Z$. The results of the \textit{ab initio} calculations of the QED corrections are compared with those obtained with the approximate effective operators \cite{Hegstrom:73:pra,glazov:04:pra} based on the free-electron anomalous magnetic moment. For the $2p_{1/2}$ state a reasonable agreement for low $Z$ ions was found, whereas for the $s$ states it was concluded that this scheme with effective operators is incomplete. The developed methods will be used to calculate the QED corrections to the second-order Zeeman splitting in lithium- and boron-like ions, taking into account the interelectronic-interaction effects.

%
%
\section*{Acknowledgements}

This work was supported by the Ministry of Science and Higher Education of the Russian Federation (Project No. FSER-2025–0012) and by the Russian Science Foundation (Project No. 23-62-10026).
The work of E.~A.~P. and A.~V.~M. was supported by the Foundation for the Advancement of Theoretical Physics and Mathematics BASIS (Project No. 24-1-2-74-1).
The work of D.~A.~G. was supported by the Foundation for the Advancement of Theoretical Physics and Mathematics BASIS (Project No. 23-1-2-52-1).
We also thank Ivan S. Terekhov for enlightening and encouraging discussions on the subject.

%
%
%

%
\appendix
\section{Self-energy ultraviolet divergences \label{sec:app:1}}

In the present Appendix, we discuss the general cancellation scheme of the ultraviolet (UV) divergences in the self-energy correction to the quadratic Zeeman effect.

The total self-energy correction should be UV finite.  According to the conventional renormalization procedure, see, e.g., Ref.~\cite{Snyderman:1991:AP}, each self-energy diagram  must be accompanied by the corresponding mass-counterterm diagram. Therefore $\Sigma(\varepsilon_a)$, defined in Eq.~(\ref{eq:oper-beg}), has to be replaced with 
\begin{equation}
\label{eq:regularization}
    \Sigma_R(\varepsilon) \equiv \Sigma(\varepsilon) - \beta \delta m
\,.
\end{equation}
However, the matrix elements of $\Sigma_R$ still have UV divergence. Restricting the consideration to the matrix elements for the states \ket{q} and \ket{p} that obey the Dirac  equation (\ref{eq:dirac}), these divergences can be parametrized as follows \cite{Glazov:10:pra},
\begin{equation}
\label{eq:uv1}
    \langle p|\Sigma_R(\varepsilon)|q\rangle[\textrm{UV}]= B^{(1)}\delta_{pq}(\varepsilon-\varepsilon_p)
\,,
\end{equation}
where $B^{(1)}$ is the UV-divergent constant.
The matrix elements of the vertex operator (\ref{eq:gumo}) also have divergences,
\begin{equation}
\label{eq:uv2}
    \langle p|\Gamma [\Umag](\varepsilon)|q\rangle[\textrm{UV}] = L^{(1)} \langle p|\Umag|q\rangle,
\end{equation}
where $L^{(1)}$ is the corresponding UV-divergent constant. Using the Ward identity, one can derive the following relation between the constants: $L^{(1)}=-B^{(1)}$. From Eq.~(\ref{eq:uv1}), it can be seen that the divergence of $\Sigma_R$ is linear in the energy parameter $\varepsilon$. Therefore, one straightforwardly obtains,
\begin{equation}
\label{eq:uv3}
    \langle p|\Sigma_R'(\varepsilon)|q\rangle[\textrm{UV}]= B^{(1)}\delta_{pq},
\end{equation}
while the matrix elements of $\Sigma''_R$ are UV-finite. According to Eq. (\ref{eq:uv2}), the matrix elements of the vertex operator $\Gamma[\Umag]$ are independent of $\varepsilon$. Consequently, the derivation over $\varepsilon$ makes them UV finite also. Finally, a simple count of vertices in the operator $\Lambda[\Umag,\Umag]$ convinces us that the corresponding expression does not contain any UV divergence.

Employing Eqs. (\ref{eq:uv1})-(\ref{eq:uv3}) and the formal expressions for the self-energy correction, given in Eqs. (\ref{eq:se:start})-(\ref{eq:se:end}), we find that
\begin{align}
\label{eq:uv1:B}
    \dgqse{B}{}[\textrm{UV}] &
        = \sideset{}{'} \sum_{n_1,n_2}\frac{\langle a|\Umag|n_1\rangle B^{(1)}\delta_{n_1n_2}(\varepsilon_a-\varepsilon_{n_1})\langle n_2|\Umag|a\rangle}{(\varepsilon_a-\varepsilon_{n_1})(\varepsilon_a-\varepsilon_{n_2})}
        = B^{(1)}\sumnn\frac{\langle a|\Umag|n\rangle\langle n|\Umag|a\rangle}{\depsn{}}
\,,\\
\label{eq:uv1:C}
    \dgqse{C}{}[\textrm{UV}]&=2L^{(1)}\sumnn\frac{\langle a|\Umag|n\rangle\langle n|\Umag|a\rangle}{\depsn{}}
\,,\\
\label{eq:uv1:H2}
    \dgqse{H2}{}[\textrm{UV}]&=B^{(1)}\sumnn\frac{\langle a|\Umag|n\rangle\langle n|\Umag|a\rangle}{\depsn{}}
\,,
\end{align}
while all the other contributions $\dgqse{X}{}$ with $X\in\{A,D,G1,G2,H1,I1,I2\}$ are UV-finite, i.e., $\dgqse{X}{}[\textrm{UV}]=0$ for those $X$.
We see that the ultraviolet divergences of the self-energy correction cancel out,
\begin{equation}
    \dgqse{\mathrm{SE}}{}[\textrm{UV}]=2(B^{(1)}+L^{(1)})\sumnn\frac{\langle a|\Umag|n\rangle\langle n|\Umag|a\rangle}{\depsn{}}=0
\,.
\end{equation}

\section{Self-energy infrared divergences \label{sec:app:irr}}
\label{ap:ir}

The individual self-energy terms, namely, $D$, $H2$, $I1$, and $I2$, exhibit infrared-divergent (IR) behavior of several types. IR divergences typically occur when the intermediate-state energies in the self-energy loops coincide with the reference-state energy $\varepsilon_a$, and are thus associated with the denominators of the form 
$P_s(\omega) \equiv 1/(-\omega+i0)^s$
with $s=2,3,\ldots$. The IR divergences disappear when the corresponding contributions are regularized in a similar way and evaluated together. In the present Appendix, we demonstrate the cancellation of the IR divergences for the self-energy correction to the quadratic Zeeman effect. The related finite residuals are briefly discussed as well. Note, however, that in practical calculations we prefer to treat the IR divergences numerically by combining together all the relevant contributions before evaluating the integrations over $\omega$. The method employed to treat the IR divergences follows the procedure discussed, e.g., in Ref.~\cite{yerokhin:10:pra}, where the case $s=2$ was considered in the context of the self-energy correction to the $g$ factor or hyperfine splitting, see also Ref.~\cite{Yerokhin:2020:800}.

First, we study the master integrals that arise when studying IR divergences. For simplicity, we work in the Feynman gauge, in which the operator $I(\omega,r_{12})$~(\ref{eq:I_omega}) for a finite photon mass $\mu$ (not to be confused with a Lorentzian index) is conveniently given by
\begin{align}
\label{eq:app:I}
    I_\mu(\omega,r_{12}) =  -e^2 \alpha_\nu\alpha^\nu \int 
    \! \frac{d{\bm k}}{(2\pi)^3} \frac{\exp(i{\bm k}\cdot{\bm r}_{12})}{\omega^2-{\bm k}^2-\mu^2+i0} 
\,.
\end{align}
Using Eq.~(\ref{eq:app:I}), the desired master integrals can be written as
\begin{align}
\label{eq:appC:Js}
    J_s(\mu,r_{12}) = \frac{i}{2\pi} \int\limits_{-\infty}^\infty \! d\omega \,
        \frac{I_\mu(\omega,r_{12})}{(-\omega+i0)^s} 
        \equiv \alpha\, \alpha_\nu\alpha^\nu \, \tilde{J}_s(\mu,r_{12})
\,.
\end{align}
We transform the operator $\tilde{J}_s$, defined by Eq.~(\ref{eq:appC:Js}), by successively applying the following manipulations. First, we evaluate the integral over $\omega$ by closing the integration contour in the lower half-plane and using the Cauchy's theorem. Second, we evaluate the angular integrals and perform the $k$-integration by parts. The result is
\begin{align}
\label{eq:app:J_s_fin}
    \tilde{J}_s(\mu,r_{12}) 
        = (-1)^{s+1} \! \int \! \frac{d{\bm k}}{(2\pi)^2} \frac{\exp(i{\bm k}\cdot{\bm r}_{12})}{\left(\sqrt{{\bm k}^2+\mu^2}\right)^{s+1}}
        = \frac{(-1)^{s+1}}{(s-1)\pi} \, \int\limits_0^\infty \! dk \, \frac{\cos(kr_{12})}{\left(\sqrt{k^2+\mu^2}\right)^{s-1}} 
\,.
\end{align}
By adding and subtracting $\cos k$ in the numerator of 
(\ref{eq:app:J_s_fin}), we represent $\tilde{J}_s$ as the sum of two terms: the first one, $\tilde{J}^{(c)}_s$, is convergent as $\mu\rightarrow 0$, while the second, $\tilde{J}^{(d)}_s$, is divergent but, crucially, does not depend on $r_{12}$,
\begin{align}
\label{eq:appC:Js:c}
    \tilde{J}^{(c)}_s(\mu,r_{12}) &\equiv \frac{(-1)^{s+1}}{(s-1)\pi} \, \int\limits_0^\infty \! dk \, \frac{\cos(kr_{12})-\cos k}{\left(\sqrt{k^2+\mu^2}\right)^{s-1}} 
\,,\\
\label{eq:appC:Js:d}
    \tilde{J}^{(d)}_s(\mu) &\equiv \frac{(-1)^{s+1}}{(s-1)\pi} \, \int\limits_0^\infty \! dk \, \frac{\cos k}{\left(\sqrt{k^2+\mu^2}\right)^{s-1}} 
\,.
\end{align}
For the matrix element of the divergent part of Eq.~(\ref{eq:appC:Js}), $J^{(d)}_s$, one obtains
\begin{align}
\label{eq:appC:Js:c:matr}
    \langle p_1 p_2 | J^{(d)}_s(\mu) | q_1 q_2 \rangle
        = \alpha \left[ \delta_{p_1q_1}\delta_{p_2q_2} - \langle p_1 | \balpha | q_1 \rangle \cdot \langle p_2 | \balpha | q_2 \rangle \right ] \tilde{J}^{(d)}_s(\mu)  
\,. 
\end{align} 
When analyzing the IR behavior, we assume that $|p_i\rangle$ and $|q_i\rangle$ are solutions of the one-electron Dirac equation~(\ref{eq:dirac}). Then, the following formula is of great practical use:
\begin{align}
\label{eq:app:formula}
    \langle p | \balpha | q \rangle =
        i \langle p | [  h_{\rm D}, {\bm r} ] | q \rangle =
        i ( \varepsilon_p - \varepsilon_q ) \langle p | {\bm r} | q \rangle 
\,.
\end{align}
In particular, it shows that the states of the same energy and the opposite parity compared to the reference state, e.g., $|a\rangle=2s$ and $|\tilde{a}\rangle=2p_{1/2}$ in the case of a point nucleus, do not cause any IR divergence, since $\langle a|\tilde{a}\rangle=0$ due to the orthogonality and $\langle a | \balpha | \tilde{a} \rangle =0$ according to Eq.~(\ref{eq:app:formula}). 
Assuming that $|a'\rangle$, $|a''\rangle$, \dots denote the intermediate states that are degenerate in energy with the reference state $|a\rangle$ and have the same parity and different total angular-momentum projections, one derives from Eq.~(\ref{eq:appC:Js:c:matr}),
\begin{align}
\label{eq:appC:Js:c:matr:aaaa}
    \langle a a' | J^{(d)}_s(\mu) | a'' a''' \rangle = \alpha \, \delta_{M_{a}M_{a''}}\delta_{M_{a'}M_{a'''}} \,   \tilde{J}^{(d)}_s(\mu) 
\,.
\end{align}
For any $|n\rangle$ that differs from $|a\rangle$ in energy, $\varepsilon_n \neq \varepsilon_a$, one finds,
\begin{align}
\label{eq:appC:Js:c:matr:aaan}
    \langle a a' | J^{(d)}_s(\mu) | a'' n \rangle = 0 
\,,
\end{align}
because $\langle a |\balpha|a''\rangle =0$ and $\delta_{a'n}=0$. Index permutations lead to an obvious generalization of the formulas (\ref{eq:appC:Js:c:matr:aaaa}) and (\ref{eq:appC:Js:c:matr:aaan}).
For the specific cases $s=2$ and $s=3$, one can easily obtain,
\begin{eqnarray}
\label{eq:app:J2}
    \tilde{J}^{(c)}_2(0,r_{12}) &\!= \dfrac{1}{\pi} {\rm ln\,}r_{12} 
\, , \qquad
    \tilde{J}^{(d)}_2(\mu) &\!= \dfrac{1}{\pi}  
        {\rm ln  \,}\dfrac{\mu}{2} + \dfrac{\gamma_E}{\pi}   + O(\mu) 
\, , \\
\label{eq:app:J3} 
    \tilde{J}^{(c)}_3(0,r_{12}) &\!= \dfrac{1}{4} - \dfrac{r_{12}}{4} 
\, , \qquad
        \tilde{J}^{(d)}_3(\mu) &\!= \dfrac{1}{4\mu} - \dfrac{1}{4} + O(\mu) 
\, ,
\end{eqnarray}
where $\gamma_E$ is the  Euler's constant.

Now we proceed to the cancellation of the IR divergences for the self-energy correction to the quadratic Zeeman effect.
The terms $\dgqse{X}{}$ with $X\in\{A,B,G1,G2\}$ are obviously IR finite, since there are no coinciding denominators in them. Let us carefully analyze all the remaining contributions by introducing the finite photon mass $\mu$.

In $\dgqse{C}{}$, the denominator $1/(-\omega+i0)^2$ occurs in the vertex operator, when $|n_1\rangle=|a'\rangle$ and $|n_2\rangle=|a''\rangle$. The term suspicious on the IR behavior reads as
\begin{align}
\label{eq:app:C_IR}
    \dgqse{C}{}[{\rm IR}]=
        2 \, \sumnn \! \sum_{M_{a'}, M_{a''}} \frac{\langle a a'' | J^{(d)}_2(\mu) |a' n\rangle 
        \langle a' |\Umag|a''\rangle \langle n | \Umag | a \rangle }{\varepsilon_a-\varepsilon_n} =0
\,,
\end{align}
where Eq.~(\ref{eq:appC:Js:c:matr:aaan}) is used.
Therefore, $\dgqse{C}{}$ is IR finite.

In the contribution $\dgqse{D}{}$, the second-order pole $P_2(\omega)=1/(-\omega+i0)^2$ arises in three cases: (i) $\varepsilon_{n_1}=\varepsilon_{n_2}=\varepsilon_{a}$, $\varepsilon_{n_3}\neq \varepsilon_{a}$; (ii) $\varepsilon_{n_1}=\varepsilon_{n_3}=\varepsilon_{a}$, $\varepsilon_{n_2}\neq \varepsilon_{a}$; (iii) $\varepsilon_{n_2}=\varepsilon_{n_3}=\varepsilon_{a}$, $\varepsilon_{n_1}\neq \varepsilon_{a}$. The first and third cases are identical. Considering them together, we obtain
\begin{align}
\label{eq:app:D_i_iii}
    \dgqse{D}{}[P_2(\omega)]_{{\rm(i)+(iii)}}=
        2 \,\frac{i}{2\pi} \int\limits_{-\infty}^\infty \! d\omega \, \sum_n^{\varepsilon_n\neq \varepsilon_a} \! \sum_{M_{a'}, M_{a''}}
        \frac{\langle a n | I_\mu(\omega) | a' a \rangle \langle a' | \Umag | a'' \rangle \langle a'' | \Umag | n \rangle}{(-\omega+i0)^2(\varepsilon_a-\omega-\varepsilon_{n}^-)}
\, .
\end{align}
Let us transform the denominator $1/(\varepsilon_a-\omega-\varepsilon_{n}^-)$ by adding and subtracting $1/(\varepsilon_a-\varepsilon_n)$,
\begin{align}
\label{eq:app:denom}
\frac{1}{\varepsilon_a-\omega-\varepsilon_{n}^-}=
\frac{\omega}{(\varepsilon_a-\omega-\varepsilon_{n}^-)(\varepsilon_a-\varepsilon_n)}
+
\frac{1}{\varepsilon_a-\varepsilon_n}
\end{align}
Here $\omega$ in the numerator reduces the order of the pole at $\omega=0$ in Eq.~(\ref{eq:app:D_i_iii}), so the first term obviously does not lead to the IR divergence. Substituting the second term in Eq.~(\ref{eq:app:denom}) into Eq.~(\ref{eq:app:D_i_iii}), one finds,
\begin{align}
\label{eq:app:D_i_iii:IR}
    \dgqse{D}{}[{\rm IR\text{-2}}]_{{\rm(i)+(iii)}}=
        2 \, \sum_n^{\varepsilon_n\neq \varepsilon_a} \! \sum_{M_{a'}, M_{a''}}
        \frac{\langle a n | J^{(d)}_2(\mu) | a' a \rangle \langle a' | \Umag | a'' \rangle \langle a'' | \Umag | n \rangle }{\varepsilon_a - \varepsilon_n} =0 
\,.
\end{align}
Therefore, the cases (i) and (iii) are IR finite. For the case (ii), we obtain
\begin{align}
\label{eq:app:D_ii}
    \dgqse{D}{}[P_2(\omega)]_{{\rm(ii)}}=
        \frac{i}{2\pi} \int\limits_{-\infty}^\infty \! d\omega \, \sum_n^{\varepsilon_n\neq \varepsilon_a} \! \sum_{M_{a'}, M_{a''}}
        \frac{\langle a a'' | I_\mu(\omega) | a' a \rangle \langle a' | \Umag | n \rangle \langle n | \Umag | a'' \rangle}{(-\omega+i0)^2(\varepsilon_a-\omega-\varepsilon_{n}^-)}
\,.
\end{align}
Application of the identity (\ref{eq:app:denom}) results in
\begin{align}
\label{eq:app:D:IR_2}
    \dgqse{D}{}[{\rm IR\text{-2}}]_{{\rm(ii)}}&=
        \sum_n^{\varepsilon_n\neq \varepsilon_a} \! \sum_{M_{a'}, M_{a''}}
        \frac{\langle a a'' | J^{(d)}_2(\mu) | a' a \rangle \langle a' | \Umag | n \rangle \langle n | \Umag | a'' \rangle }{\varepsilon_a - \varepsilon_n} 
\nonumber \\
    &= \alpha \tilde{J}^{(d)}_2(\mu)  \sumnn \! \frac{\langle a | \Umag | n \rangle \langle n | \Umag | a \rangle }{\varepsilon_a - \varepsilon_n} 
\,,
\end{align}
where Eq.~(\ref{eq:appC:Js:c:matr:aaaa}) is employed.
Third-order pole $P_3(\omega)=1/(-\omega+i0)^3$ also arises in $\dgqse{D}{}$,
\begin{align}
\label{eq:app:D:IR_3}
    \dgqse{D}{}[{\rm IR\text{-3}}] &=\!\!\!\!\!
        \sum_{M_{a'}, M_{a''}, M_{a'''}}\!\!\!\!\!
        \langle a a''' | J^{(d)}_3(\mu) | a' a \rangle \langle a' | \Umag | a'' \rangle \langle a'' | \Umag | a''' \rangle 
        =\alpha \tilde{J}^{(d)}_3(\mu) \langle a | \Umag | a \rangle^2 
\,,
\end{align}
where we have taken into account that $\Umag$ conserves the projection of the total angular momentum, $\langle p|\Umag|q\rangle\sim\delta_{M_pM_q}$.

Next, we consider the reducible contributions to the self-energy correction. In $\dgqse{H1}{}$, the denominator $1/(-\omega+i0)^2$ arises,
\begin{align}
    \dgqse{H1}{}[{\rm IR}]=
        -2 \langle a|\Umag|a\rangle \sumnn \sum_{M_{a'}}
        \frac{\langle a a' | J^{(d)}_2(\mu) | a' n \rangle \langle n | \Umag|a\rangle}{\varepsilon_a-\varepsilon_n} = 0 
\,.
\end{align}
Thus, the corresponding contribution is IR finite.
For $\dgqse{H2}{}$, we obtain,
\begin{align}
\label{eq:app:H2:IR}
    \dgqse{H2}{}[{\rm IR}]&= 
        - \sum_{M_{a'}} \langle a a' | J^{(d)}_2(\mu) | a' a \rangle 
        \sumnn \! \frac{\langle a | \Umag | n \rangle \langle n | \Umag | a \rangle }{\varepsilon_a - \varepsilon_n}  
\nonumber \\
    &= -\alpha \tilde{J}^{(d)}_2(\mu)  \sumnn \! \frac{\langle a | \Umag | n \rangle \langle n | \Umag | a \rangle }{\varepsilon_a - \varepsilon_n} 
\,.
\end{align}
For $\dgqse{I1}{}$, we have to analyze both second- and third-order poles. The term corresponding to the second-order pole $P_2(\omega)=1/(-\omega+i0)^2$ reads as
\begin{align}
    \dgqse{I1}{}[P_2(\omega)] = 
        -2 \,\frac{i}{2\pi} \int\limits_{-\infty}^\infty \! d\omega \, \sum_n^{\varepsilon_n\neq \varepsilon_a} \! \sum_{M_{a'}}
        \frac{\langle a n | I_\mu(\omega)|a' a\rangle \langle a' |\Umag | n \rangle }{(-\omega+i0)^2(\varepsilon_a-\omega-\varepsilon_{n}^-)}
        \langle a | \Umag | a \rangle 
\,.
\end{align}
Applying the identity (\ref{eq:app:denom}), we find
\begin{align}
\label{eq:app:I1:IR-2}
    \dgqse{I1}{}[{\rm IR\text{-2}}]=
        -2 \, \sum_n^{\varepsilon_n\neq \varepsilon_a} \! \sum_{M_{a'}}
        \frac{\langle a n | J^{(d)}_2(\mu) | a' a \rangle \langle a' | \Umag | n \rangle  }{\varepsilon_a - \varepsilon_n} \langle a | \Umag | a \rangle = 0
\,.
\end{align}
Therefore, the second-order pole does not exhibit the IR behavior in $\dgqse{I1}{}$. 
The third-order pole leads to
\begin{align}
\label{eq:app:I1:IR-3}
    \dgqse{I1}{}[{\rm IR\text{-3}}]&=
        -2  \sum_{M_{a'}, M_{a''}} \langle a a'' | J^{(d)}_3(\mu) | a' a \rangle \langle a' | \Umag | a'' \rangle  \langle a | \Umag | a \rangle 
\nonumber \\
    &=-2\alpha \tilde{J}^{(d)}_3(\mu) \langle a | \Umag | a \rangle^2 
\,.
\end{align}
Finally, the third-order pole arises also in $\dgqse{I2}{}$,
\begin{align}
\label{eq:app:I2:IR}
    \dgqse{I2}{}[{\rm IR}]&=
        \sum_{M_{a'}} \langle a a' | J^{(d)}_3(\mu)  | a' a \rangle \langle a | \Umag | a \rangle^2 
        = \alpha \tilde{J}^{(d)}_3(\mu) \langle a | \Umag | a \rangle^2 
\,.
\end{align}

In summary, the second-order poles $P_2(\omega)$ cause the IR divergences in the contributions $\dgqse{D}{}$ and $\dgqse{H2}{}$ (Eqs.~(\ref{eq:app:D:IR_2}) and (\ref{eq:app:H2:IR})), however, their sum is finite. The third-order poles $P_3(\omega)$ cause the IR divergences in $\dgqse{D}{}$, $\dgqse{I1}{}$, and $\dgqse{I2}{}$ (Eqs.~(\ref{eq:app:D:IR_3}), (\ref{eq:app:I1:IR-3}), and (\ref{eq:app:I2:IR})). These divergences also cancel in the sum. The finite residuals, if necessary, can be easily derived using the operators $\tilde{J}^{(c)}_2$ and $\tilde{J}^{(c)}_3$ given in Eqs.~(\ref{eq:app:J2}) and (\ref{eq:app:J3}).


%

%


\begin{table}[ht]
\caption{\label{tab:gbb}
Leading-order contribution to the quadratic Zeeman effect $\gqd{0}$ given by Eq.~(\ref{eq:gbb-dirac}) for the $1s_{1/2}$, $2s_{1/2}$, and $2p_{1/2}$ states.}
\begin{tabular}{cS[table-format=-3.5(2),group-separator=,table-align-text-post=false]S[table-format=-4.5(2),group-separator=,table-align-text-post=false]S[table-format=-6.5(2),group-separator=,table-align-text-post=false]} 
\hline
\hline
$Z$ & {$1s_{1/2}$} & {$2s_{1/2}$} & {$2p_{1/2}$} \\
\hline
14&94.47936&1330.93&-63816.12\\
16&72.02440&1016.55&-37147.38\\
18&56.62964&801.02&-23007.276\\
20&45.61813&646.844&-14960.925\\
24&31.27499&446.015&-7066.154\\
32&17.01625&246.3259&-2117.9627\\
54&5.13783&79.73767&-203.77849\\
82&1.53519&28.65524&-20.22674\\
92&0.98376&20.60414&-8.56664\\
\hline
\hline
\end{tabular}
\end{table}

\begin{table*}[t]
\centering

\caption{\label{tab:details_1s} 
         Individual contributions of the self-energy diagrams in Fig.~\ref{fig:se} to the quadratic Zeeman effect on the $1s$ state in hydrogenlike argon ($Z=18$), 
         in terms of the function $g^{(2)}$ defined in Eq.~(\ref{eq:d2_def}).
         The $|\kappa|=1$ contributions of the terms $D$, $H2$, $I1$, and $I2$ are added to the $|\kappa|=2$ contribution of the term $D$.
         The values represented by zeros are omitted for better readability.
         See the text for details.
         }
         
\resizebox{\textwidth}{!}{%
\begin{tabular}{
                c
                S[table-format=-2.5,group-separator=,table-align-text-post=false]
                S[table-format=-2.5,group-separator=,table-align-text-post=false]
                S[table-format=2.5,group-separator=,table-align-text-post=false]
                S[table-format=-2.5,group-separator=,table-align-text-post=false]
                S[table-format=-2.5,group-separator=,table-align-text-post=false]
                S[table-format=-2.5,group-separator=,table-align-text-post=false]
                S[table-format=2.5,group-separator=,table-align-text-post=false]
                S[table-format=-2.5,group-separator=,table-align-text-post=false]
                S[table-format=2.5,group-separator=,table-align-text-post=false]
                S[table-format=-2.5(2),group-separator=,table-align-text-post=false,table-space-text-post=$^{\dagger}$]
                S[table-format=-1.5(2),group-separator=,table-align-text-post=false]
               }
               
\hline
\hline
  
                                                        &      
  \multicolumn{1}{c}{$A$}        &
  \multicolumn{1}{c}{$B$}        & 
  \multicolumn{1}{c}{$C$}        &
  \multicolumn{1}{c}{$G1$}       &
  \multicolumn{1}{c}{$G2$}       &  
  \multicolumn{1}{c}{$H1$}       &
  \multicolumn{1}{c}{$H2$}       &
  \multicolumn{1}{c}{$I1$}       &
  \multicolumn{1}{c}{$I2$}       &
  \multicolumn{1}{c}{$D$}        & 
  \multicolumn{1}{c}{Sum}      \\ 
        
\hline   
                       
   Free   &         -0.16180 &         -0.05869 &          0.13335 &         -0.00206 &          0.00105 &         -0.00805 &          0.35059 & {--} & {--} & {--} &          0.25439   \rule{0pt}{3.2ex}    \\ 
   $|\kappa|=1$   &          0.18593 &         -0.00011 &          0.00147 &          0.00236 &         -0.00108 &          0.00174 &  {--}   &  {--}   &  {--}  &  {--}  &          0.19031   \\ 
  2       &          0.00307 &         -0.00030 &          0.00470 &          0.00004 &         -0.00003 &         -0.00158 &          0.00699 &         -0.01318 &          0.00621 & -0.20278$^\dagger$ &         -0.19684   \\ 
  3       &          0.00068 &         -0.00003 &          0.00084 &          0.00001 &         -0.00001 &         -0.00012 &          0.00195 &         -0.00412 &          0.00178 &         -0.05456 &         -0.05357   \\ 
  4       &          0.00020 &         -0.00001 &          0.00041 &                  &                  &         -0.00004 &          0.00083 &         -0.00179 &          0.00078 &         -0.03489 &         -0.03451   \\ 
  5       &          0.00007 &                  &          0.00025 &                  &                  &         -0.00002 &          0.00043 &         -0.00095 &          0.00041 &         -0.02504 &         -0.02485   \\ 
  6       &          0.00002 &                  &          0.00017 &                  &                  &         -0.00001 &          0.00025 &         -0.00057 &          0.00025 &         -0.01889 &         -0.01878   \\ 
  7       &                  &                  &          0.00012 &                  &                  &         -0.00001 &          0.00016 &         -0.00036 &          0.00016 &         -0.01464 &         -0.01458   \\ 
  8       &                  &                  &          0.00009 &                  &                  &                  &          0.00011 &         -0.00025 &          0.00011 &         -0.01155 &         -0.01151   \\ 
  9       &                  &                  &          0.00006 &                  &                  &                  &          0.00007 &         -0.00018 &          0.00008 &         -0.00923 &         -0.00920   \\ 
 10       &                  &                  &          0.00005 &                  &                  &                  &          0.00005 &         -0.00013 &          0.00005 &         -0.00744 &         -0.00742   \\ 
 11       &                  &                  &          0.00004 &                  &                  &                  &          0.00004 &         -0.00010 &          0.00004 &         -0.00604 &         -0.00602   \\ 
 12       &                  &                  &          0.00003 &                  &                  &                  &          0.00003 &         -0.00008 &          0.00003 &         -0.00492 &         -0.00491   \\ 
 13       &                  &                  &          0.00002 &                  &                  &                  &          0.00002 &         -0.00006 &          0.00002 &         -0.00403 &         -0.00402   \\ 
 14       &                  &                  &          0.00002 &                  &                  &                  &          0.00002 &         -0.00005 &          0.00002 &         -0.00331 &         -0.00331   \\ 
 15       &                  &                  &          0.00002 &                  &                  &                  &          0.00001 &         -0.00004 &          0.00002 &         -0.00272 &         -0.00272   \\ 
 16       &                  &                  &          0.00001 &                  &                  &                  &          0.00001 &         -0.00003 &          0.00001 &         -0.00225 &         -0.00224   \\ 
 17       &                  &                  &          0.00001 &                  &                  &                  &          0.00001 &         -0.00003 &          0.00001 &         -0.00185 &         -0.00185   \\ 
 18       &                  &                  &          0.00001 &                  &                  &                  &          0.00001 &         -0.00002 &          0.00001 &         -0.00153 &         -0.00153   \\ 
 19       &                  &                  &          0.00001 &                  &                  &                  &          0.00001 &         -0.00002 &          0.00001 &         -0.00126 &         -0.00126   \\ 
 20       &                  &                  &          0.00001 &                  &                  &                  &          0.00001 &         -0.00002 &          0.00001 &         -0.00103 &         -0.00103   \\ 
 21       &                  &                  &          0.00001 &                  &                  &                  &                  &         -0.00001 &          0.00001 &         -0.00085 &         -0.00085   \\ 
 22       &                  &                  &                  &                  &                  &                  &                  &         -0.00001 &                  &         -0.00069 &         -0.00069   \\ 
 23       &                  &                  &                  &                  &                  &                  &                  &         -0.00001 &                  &         -0.00056 &         -0.00056   \\ 
 24       &                  &                  &                  &                  &                  &                  &                  &         -0.00001 &                  &         -0.00045 &         -0.00045   \\ 
 $\sum_{|\kappa|=1}^{24}$   &          0.18993 &         -0.00047 &          0.00835 &          0.00241 &         -0.00111 &         -0.00005 &          0.01104 &         -0.02201 &          0.01002 &         -0.41053 &         -0.21242   \\ 
 $\sum_{|\kappa|\geqslant 25}$   &         -0.00001 &         -0.00000 &          0.00003 &         -0.00000 &         -0.00000 &         -0.00000 &          0.00003 &         -0.00013 &          0.00003 &      0.00432(11) &      0.00428(11)   \\ 
 Total   &          0.02813 &         -0.05915 &          0.14172 &          0.00035 &         -0.00007 &         -0.00810 &          0.36166 &         -0.02214 &          0.01005 &     -0.40621(11) &      0.04625(11)   \\ 

\hline
\hline

\end{tabular}%
}

{\bigskip
$^\dagger$ This value includes the $|\kappa|=1$ contributions of the diagrams $D$, $H2$, $I1$, and $I2$.
}

\end{table*}

\begin{table*}[t]
\centering

\caption{\label{tab:details_2s} 
         The same as in Table~\ref{tab:details_1s} for the $2s$ state.
         }
         
\resizebox{\textwidth}{!}{%
\begin{tabular}{
                c
                S[table-format=-2.5(1),group-separator=,table-align-text-post=false]
                S[table-format=-2.5,group-separator=,table-align-text-post=false]
                S[table-format=2.5(1),group-separator=,table-align-text-post=false]
                S[table-format=-2.5,group-separator=,table-align-text-post=false]
                S[table-format=-2.5,group-separator=,table-align-text-post=false]
                S[table-format=-2.5,group-separator=,table-align-text-post=false]
                S[table-format=2.5(1),group-separator=,table-align-text-post=false]
                S[table-format=-2.5(1),group-separator=,table-align-text-post=false]
                S[table-format=2.5(1),group-separator=,table-align-text-post=false]
                S[table-format=-2.2(2),group-separator=,table-align-text-post=false,table-space-text-post=$^{\dagger}$]
                S[table-format=-2.2(2),group-separator=,table-align-text-post=false]
               }
               
\hline
\hline
  
                                                        &      
  \multicolumn{1}{c}{$A$}        &
  \multicolumn{1}{c}{$B$}        & 
  \multicolumn{1}{c}{$C$}        & 
  \multicolumn{1}{c}{$G1$}       &
  \multicolumn{1}{c}{$G2$}       &  
  \multicolumn{1}{c}{$H1$}       &
  \multicolumn{1}{c}{$H2$}       &
  \multicolumn{1}{c}{$I1$}       &
  \multicolumn{1}{c}{$I2$}       &
  \multicolumn{1}{c}{$D$}        &
  \multicolumn{1}{c}{Sum}      \\ 
        
\hline   
                       
   Free   &         -2.59402 &         -0.91533 &          1.83135 &         -0.00810 &          0.00367 &         -0.01128 &          7.61178 & {--} & {--} & {--} &          5.91806   \rule{0pt}{3.2ex}    \\ 
   $|\kappa|=1$   &          2.68359 &         -0.00127 &          0.01401 &          0.00852 &         -0.00357 &          0.00120 &  {--}   &  {--}   &  {--}  &  {--}  &          2.70248   \\ 
  2       &          0.08749 &         -0.00400 &          0.01332 &          0.00021 &         -0.00013 &         -0.00139 &          0.15877 &         -0.11054 &          0.04337 & -3.43590$^\dagger$ &         -3.24880   \\ 
  3       &          0.02483 &         -0.00019 &          0.00390 &          0.00006 &         -0.00004 &         -0.00016 &          0.04569 &         -0.02713 &          0.01091 &         -0.85845 &         -0.80057   \\ 
  4       &          0.01058 &         -0.00006 &          0.00233 &          0.00002 &         -0.00002 &         -0.00005 &          0.02173 &         -0.01249 &          0.00511 &         -0.56078 &         -0.53361   \\ 
  5       &          0.00552 &         -0.00003 &          0.00166 &          0.00001 &         -0.00001 &         -0.00002 &          0.01259 &         -0.00712 &          0.00296 &         -0.42561 &         -0.41006   \\ 
  6       &          0.00323 &         -0.00002 &          0.00127 &          0.00001 &         -0.00001 &         -0.00001 &          0.00812 &         -0.00455 &          0.00191 &         -0.34457 &         -0.33462   \\ 
  7       &          0.00203 &         -0.00001 &          0.00101 &                  &                  &         -0.00001 &          0.00560 &         -0.00312 &          0.00132 &         -0.28943 &         -0.28261   \\ 
  8       &          0.00134 &         -0.00001 &          0.00083 &                  &                  &         -0.00001 &          0.00406 &         -0.00225 &          0.00096 &         -0.24901 &         -0.24409   \\ 
  9       &          0.00092 &         -0.00001 &          0.00069 &                  &                  &         -0.00001 &          0.00304 &         -0.00168 &          0.00072 &         -0.21785 &         -0.21417   \\ 
 10       &          0.00065 &         -0.00001 &          0.00059 &                  &                  &                  &          0.00234 &         -0.00129 &          0.00056 &         -0.19294 &         -0.19011   \\ 
 11       &          0.00047 &                  &          0.00050 &                  &                  &                  &          0.00185 &         -0.00102 &          0.00044 &         -0.17249 &         -0.17026   \\ 
 12       &          0.00034 &                  &          0.00044 &                  &                  &                  &          0.00148 &         -0.00082 &          0.00035 &         -0.15533 &         -0.15354   \\ 
 13       &          0.00025 &                  &          0.00038 &                  &                  &                  &          0.00120 &         -0.00067 &          0.00029 &         -0.14069 &         -0.13924   \\ 
 14       &          0.00019 &                  &          0.00034 &                  &                  &                  &          0.00099 &         -0.00055 &          0.00024 &         -0.12804 &         -0.12683   \\ 
 15       &          0.00014 &                  &          0.00030 &                  &                  &                  &          0.00083 &         -0.00046 &          0.00020 &         -0.11697 &         -0.11597   \\ 
 16       &          0.00010 &                  &          0.00026 &                  &                  &                  &          0.00070 &         -0.00039 &          0.00017 &         -0.10720 &         -0.10636   \\ 
 17       &          0.00008 &                  &          0.00024 &                  &                  &                  &          0.00059 &         -0.00033 &          0.00014 &         -0.09852 &         -0.09780   \\ 
 18       &          0.00006 &                  &          0.00021 &                  &                  &                  &          0.00051 &         -0.00028 &          0.00012 &         -0.09075 &         -0.09013   \\ 
 19       &          0.00004 &                  &          0.00019 &                  &                  &                  &          0.00044 &         -0.00024 &          0.00011 &         -0.08376 &         -0.08322   \\ 
 20       &          0.00003 &                  &          0.00017 &                  &                  &                  &          0.00038 &         -0.00021 &          0.00009 &         -0.07743 &         -0.07697   \\ 
 21       &          0.00002 &                  &          0.00016 &                  &                  &                  &          0.00033 &         -0.00019 &          0.00008 &         -0.07169 &         -0.07129   \\ 
 22       &          0.00002 &                  &          0.00014 &                  &                  &                  &          0.00029 &         -0.00016 &          0.00007 &         -0.06646 &         -0.06611   \\ 
 23       &          0.00001 &                  &          0.00013 &                  &                  &                  &          0.00025 &         -0.00014 &          0.00006 &         -0.06168 &         -0.06137   \\ 
 24       &          0.00001 &                  &          0.00012 &                  &                  &                  &          0.00022 &         -0.00013 &          0.00006 &         -0.05730 &         -0.05703   \\ 
 $\sum_{|\kappa|=1}^{24}$   &          2.82194 &         -0.00562 &          0.04318 &          0.00885 &         -0.00380 &         -0.00048 &          0.27201 &         -0.17576 &          0.07025 &         -8.00285 &         -4.97228   \\ 
 $\sum_{|\kappa|\geqslant 25}$   &      -0.00018(2) &         -0.00001 &       0.00157(5) &         -0.00000 &         -0.00000 &         -0.00001 &       0.00246(1) &      -0.00157(2) &       0.00063(1) &        -0.72(14) &        -0.72(14)   \\ 
 Total   &       0.22774(2) &         -0.92096 &       1.87610(5) &          0.00075 &         -0.00013 &         -0.01177 &       7.88625(1) &      -0.17733(2) &       0.07088(1) &        -8.73(14) &         0.23(14)   \\ 

\hline
\hline

\end{tabular}%
}

{\bigskip
$^\dagger$ This value includes the $|\kappa|=1$ contributions of the diagrams $D$, $H2$, $I1$, and $I2$.
}

\end{table*}

\begin{table*}[t]
\centering

\caption{\label{tab:details_2p1}
         The same as in Table~\ref{tab:details_1s} for the $2p_{1/2}$ state.
         }
         
\resizebox{\textwidth}{!}{%
\begin{tabular}{
                c
                S[table-format=-2.5(1),group-separator=,table-align-text-post=false]
                S[table-format=-6.3(2),group-separator=,table-align-text-post=false]
                S[table-format=5.4(2),group-separator=,table-align-text-post=false]
                S[table-format=-2.5,group-separator=,table-align-text-post=false]
                S[table-format=-6.3(2),group-separator=,table-align-text-post=false]
                S[table-format=-2.5,group-separator=,table-align-text-post=false]
                S[table-format=-4.5(2),group-separator=,table-align-text-post=false]
                S[table-format=-2.5,group-separator=,table-align-text-post=false]
                S[table-format=2.5,group-separator=,table-align-text-post=false]
                S[table-format=-4.3(2),group-separator=,table-align-text-post=false,table-space-text-post=$^{\dagger}$]
                S[table-format=-5.3(2),group-separator=,table-align-text-post=false]
               }
               
\hline
\hline
  
                                                        &      
  \multicolumn{1}{c}{$A$}        &
  \multicolumn{1}{c}{$B$}        & 
  \multicolumn{1}{c}{$C$}        & 
  \multicolumn{1}{c}{$G1$}       &
  \multicolumn{1}{c}{$G2$}       &  
  \multicolumn{1}{c}{$H1$}       &
  \multicolumn{1}{c}{$H2$}       &
  \multicolumn{1}{c}{$I1$}       &
  \multicolumn{1}{c}{$I2$}       &
  \multicolumn{1}{c}{$D$}        &
  \multicolumn{1}{c}{Sum}      \\ 
        
\hline   
                       
   Free   &         -2.39124 &     -24976.90715 &        320.02660 &         -0.00052 &      24968.19336 &         -0.00152 &       -208.05147 & {--} & {--} & {--} &        100.86807   \rule{0pt}{3.2ex}    \\ 
   $|\kappa|=1$   &          1.28251 &        604.08026 &         10.87052 &          0.00050 &     -23383.75811 &          0.00067 &  {--}   &  {--}   &  {--}  &  {--}  &     -22767.52366   \\ 
  2       &          0.04425 &      22536.13105 &        745.62466 &          0.00002 &       -901.02671 &         -0.00001 &         -4.10286 &         -0.01051 &          0.00329 & -377.98425$^\dagger$ &      21998.67894   \\ 
  3       &          0.01393 &        743.53502 &          5.77341 &          0.00001 &       -273.23820 &                  &         -1.23987 &         -0.00289 &          0.00094 &         -0.62021 &        474.22213   \\ 
  4       &          0.00626 &        214.14497 &          1.68026 &                  &       -128.74073 &                  &         -0.59037 &         -0.00136 &          0.00044 &         -0.40355 &         86.09592   \\ 
  5       &          0.00334 &        100.09556 &          0.78875 &                  &        -73.18942 &                  &         -0.33995 &         -0.00079 &          0.00026 &         -0.30482 &         27.05293   \\ 
  6       &          0.00196 &         57.25478 &          0.45694 &                  &        -46.20075 &                  &         -0.21737 &         -0.00050 &          0.00016 &         -0.24553 &         11.04969   \\ 
  7       &          0.00123 &         36.52648 &          0.29704 &                  &        -31.20972 &                  &         -0.14865 &         -0.00035 &          0.00011 &         -0.20514 &          5.26100   \\ 
  8       &          0.00080 &         24.96323 &          0.20779 &                  &        -22.11997 &                  &         -0.10659 &         -0.00025 &          0.00008 &         -0.17548 &          2.76961   \\ 
  9       &          0.00054 &         17.89610 &          0.15301 &                  &        -16.25381 &                  &         -0.07918 &         -0.00018 &          0.00006 &         -0.15260 &          1.56394   \\ 
 10       &          0.00037 &         13.29247 &          0.11707 &                  &        -12.28619 &                  &         -0.06046 &         -0.00014 &          0.00005 &         -0.13431 &          0.92885   \\ 
 11       &          0.00026 &         10.14842 &          0.09227 &                  &         -9.50220 &                  &         -0.04721 &         -0.00011 &          0.00004 &         -0.11930 &          0.57216   \\ 
 12       &          0.00018 &          7.92118 &          0.07448 &                  &         -7.48993 &                  &         -0.03755 &         -0.00009 &          0.00003 &         -0.10672 &          0.36159   \\ 
 13       &          0.00013 &          6.29654 &          0.06130 &                  &         -5.99933 &                  &         -0.03032 &         -0.00007 &          0.00002 &         -0.09602 &          0.23226   \\ 
 14       &          0.00009 &          5.08263 &          0.05129 &                  &         -4.87211 &                  &         -0.02482 &         -0.00006 &          0.00002 &         -0.08679 &          0.15026   \\ 
 15       &          0.00006 &          4.15715 &          0.04352 &                  &         -4.00446 &                  &         -0.02055 &         -0.00005 &          0.00002 &         -0.07874 &          0.09695   \\ 
 16       &          0.00005 &          3.43936 &          0.03736 &                  &         -3.32629 &                  &         -0.01718 &         -0.00004 &          0.00001 &         -0.07168 &          0.06159   \\ 
 17       &          0.00003 &          2.87431 &          0.03241 &                  &         -2.78904 &                  &         -0.01450 &         -0.00003 &          0.00001 &         -0.06542 &          0.03777   \\ 
 18       &          0.00002 &          2.42369 &          0.02836 &                  &         -2.35832 &                  &         -0.01233 &         -0.00003 &          0.00001 &         -0.05985 &          0.02156   \\ 
 19       &          0.00001 &          2.06017 &          0.02502 &                  &         -2.00932 &                  &         -0.01056 &         -0.00002 &          0.00001 &         -0.05486 &          0.01045   \\ 
 20       &          0.00001 &          1.76390 &          0.02223 &                  &         -1.72382 &                  &         -0.00911 &         -0.00002 &          0.00001 &         -0.05038 &          0.00282   \\ 
 21       &                  &          1.52021 &          0.01987 &                  &         -1.48823 &                  &         -0.00790 &         -0.00002 &          0.00001 &         -0.04633 &         -0.00239   \\ 
 22       &                  &          1.31808 &          0.01786 &                  &         -1.29228 &                  &         -0.00689 &         -0.00001 &          0.00001 &         -0.04267 &         -0.00591   \\ 
 23       &                  &          1.14915 &          0.01613 &                  &         -1.12813 &                  &         -0.00604 &         -0.00001 &                  &         -0.03934 &         -0.00824   \\ 
 24       &                  &          1.00699 &          0.01464 &                  &         -0.98971 &                  &         -0.00532 &         -0.00001 &                  &         -0.03631 &         -0.00972   \\ 
 $\sum_{|\kappa|=1}^{24}$   &          1.35602 &      24399.08171 &        766.50618 &          0.00053 &     -24936.99678 &          0.00065 &         -7.13559 &         -0.01753 &          0.00560 &       -381.18029 &       -158.37951   \\ 
 $\sum_{|\kappa|\geqslant 25}$   &      -0.00017(1) &       10.435(17) &       0.2421(11) &         -0.00000 &      -10.324(18) &         -0.00000 &     -0.05679(20) &         -0.00001 &          0.00005 &       -0.405(77) &       -0.108(80)   \\ 
 Total   &      -1.03539(1) &     -567.390(17) &    1086.7749(11) &          0.00000 &       20.873(18) &         -0.00087 &   -215.24384(20) &         -0.01754 &          0.00565 &     -381.585(77) &      -57.620(80)   \\ 

\hline
\hline

\end{tabular}%
}

{\bigskip
$^\dagger$ This value includes the $|\kappa|=1$ contributions of the diagrams $D$, $H2$, $I1$, and $I2$.
}

\end{table*}

{\renewcommand{\arraystretch}{0.9}
\begin{table}[ht]
\caption{\label{tab:qed} 
The quantum-electrodynamical corrections, self-energy (SE) and vacuum polarization (VP), to the quadratic Zeeman effect 
for the $1s_{1/2}$, $2s_{1/2}$, and $2p_{1/2}$ states 
in terms of the function $g^{(2)}$ defined in Eq.~(\ref{eq:d2_def}). 
For $2p_{1/2}$ state, the full QED correction is also given in terms of the function $\faz$ 
defined in (\ref{eq:faz}).
         }
\begin{tabular}{ccS[table-format=-1.6(2),group-separator=,table-align-text-post=false]S[table-format=-3.6(2),group-separator=,table-align-text-post=false]S[table-format=-1.6(2),group-separator=,table-align-text-post=false]S[table-format=-1.5(2),group-separator=,table-align-text-post=false]} 
\hline
\hline
$Z$ & Term & {$1s_{1/2}$} & {$2s_{1/2}$} & \multicolumn{2}{c}{$2p_{1/2}$} \\
 & &$ \gbb$& $\gbb$& $\gbb$&{$\faz$}\\
\hline
14&SE&0.05358(64)&0.15(36)&-155.84(23)&\\
&VP&-0.00104(0)&-0.01(0)&0.32(0)&\\
&QED&0.05255(64)&0.15(36)&-155.53(23)&1.0492(15)\\
\hline
16&SE&0.04965(22)&0.21(23)&-91.82(13)&\\
&VP&-0.00103(0)&-0.01(0)&0.24(0)&\\
&QED&0.04863(22)&0.20(23)&-91.58(13)&1.0613(15)\\
\hline
18&SE&0.04625(11)&0.23(14)&-57.620(80)&\\
&VP&-0.00102(0)&-0.01(0)&0.193(0)&\\
&QED&0.04523(11)&0.22(14)&-57.427(80)&1.0746(15)\\
\hline
20&SE&0.04327(8)&0.228(86)&-37.999(47)&\\
&VP&-0.00101(0)&-0.008(0)&0.158(0)&\\
&QED&0.04226(8)&0.221(86)&-37.841(47)&1.0889(14)\\
\hline
24&SE&0.03826(5)&0.217(33)&-18.510(17)&\\
&VP&-0.00099(0)&-0.008(0)&0.111(0)&\\
&QED&0.03727(5)&0.210(33)&-18.399(17)&1.1210(11)\\
\hline
32&SE&0.03081(2)&0.1865(50)&-5.9636(27)&\\
&VP&-0.00097(0)&-0.0075(0)&0.0649(0)&\\
&QED&0.02984(2)&0.1791(50)&-5.8987(27)&1.19901(54)\\
\hline
54&SE&0.01878(2)&0.13230(30)&-0.75358(27)&\\
&VP&-0.00096(0)&-0.00775(0)&0.02635(0)&\\
&QED&0.01783(2)&0.12455(30)&-0.72723(27)&1.53638(58)\\
\hline
82&SE&0.01081(0)&0.09827(6)&-0.13229(3)&\\
&VP&-0.00101(0)&-0.00922(0)&0.01449(0)&\\
&QED&0.00981(0)&0.08905(6)&-0.11780(3)&2.5072(6)\\
\hline
92&SE&0.00888(0)&0.09025(3)&-0.07738(2)&\\
&VP&-0.00104(0)&-0.01017(0)&0.01258(0)&\\
&QED&0.00784(0)&0.08008(3)&-0.06480(2)&3.2562(8)\\
\hline
\hline
\end{tabular}
\end{table}
}
\begin{table}[ht]
\caption{\label{tab:qed-nonrel} 
Approximate treatment of the quadratic Zeeman splitting for the $2p_{1/2}$ state, based on the effective operators $\ha$ and $\hb$. The contributions corresponding to both operators as well as their sum are shown. The total QED correction from Table \ref{tab:qed} obtained by means of the \textit{ab initio} approach is given for comparison. The difference between the rigorous and approximate values, as well as its ratio to the QED correction are also given.}
\resizebox{\textwidth}{!}{%
\begin{tabular}{cS[table-format=-4.5(2),group-separator=,table-align-text-post=false]S[table-format=-1.5(2),group-separator=,table-align-text-post=false]S[table-format=-3.5(2),group-separator=,table-align-text-post=false]S[table-format=-1.5(2),group-separator=,table-align-text-post=false]S[table-format=-3.5(2),group-separator=,table-align-text-post=false]S[table-format=-1.5(2),group-separator=,table-align-text-post=false]} 
\hline
\hline
{$Z$} & {$\ha$} & {$\hb$} & {$\ha + \hb$} & $\mathrm{QED}$ & {$\Delta = \mathrm{QED} - (\ha + \hb)$} & $\Delta / \mathrm{QED}$\\[2pt]
\hline
14&-300.80(0)&151.59(0)&-149.22(0)&-155.53(23)&-6.31(23)&0.0406\\
16&-175.90(0)&88.86(0)&-87.04(0)&-91.58(13)&-4.54(13)&0.0496\\
18&-109.513(0)&55.479(0)&-54.033(0)&-57.427(80)&-3.394(79)&0.0591\\
20&-71.630(0)&36.403(0)&-35.227(0)&-37.841(47)&-2.614(47)&0.0691\\
24&-34.296(0)&17.560(0)&-16.736(0)&-18.399(17)&-1.663(17)&0.0904\\
32&-10.6506(0)&5.5619(0)&-5.0888(0)&-5.8987(27)&-0.8100(27)&0.1373\\
54&-1.20723(0)&0.69470(0)&-0.51253(0)&-0.72723(27)&-0.21470(27)&0.2952\\
82&-0.18528(0)&0.14194(0)&-0.04334(0)&-0.11779(3)&-0.07445(3)&0.6321\\
92&-0.10450(0)&0.09600(0)&-0.00850(0)&-0.06480(2)&-0.05629(2)&0.8688\\
\hline
\hline
\end{tabular}
}
\end{table}

\end{document}